\documentclass[final,1p,times,number]{elsarticle}

\usepackage{amsmath}
\usepackage{amssymb}
\usepackage{amsfonts}
\usepackage{graphicx}
\usepackage{epstopdf}
\usepackage{color}
\usepackage{bm}
\usepackage{multirow}
\usepackage{natbib}
\usepackage{hyperref}

\usepackage{ifpdf}
\ifpdf%
\usepackage{pdflscape}
\else
\usepackage{lscape}
\fi

\DeclareMathOperator{\Tr}{Tr}

\DeclareMathOperator{\sgn}{sgn}

\DeclareMathOperator{\im}{Im}

\DeclareMathOperator{\re}{Re}

\journal{Annals of Physics}
\biboptions{sort&compress}

\begin{document}

\begin{frontmatter}
\title{The effect of superconducting fluctuations on the ac conductivity of a 2D electron system in the diffusive regime}

\author[LITP]{I. S. Burmistrov\corref{cor}}

\address[LITP]{L. D. Landau Institute for Theoretical Physics, Semenova 1a, 142432, Chernogolovka, Russia}
\cortext[cor]{Corresponding author. Fax: \texttt{+7-495-702-9317} e-mail: burmi@itp.ac.ru}

\begin{abstract}
We report a complete analytical expression for the one-loop correction to the ac conductivity $\sigma(\omega)$ of a disordered two-dimensional electron system in the diffusive regime. The obtained expression includes the weak localization and Altshuler--Aronov corrections as well as the corrections due to superconducting fluctuations above superconducting transition temperature. The derived expression has no $1/(i\omega)$ divergency in the static limit, $\omega\to 0$, in agreement with general expectations for the normal state conductivity of a disordered electron system. 
\end{abstract}


\end{frontmatter}

\section{Introduction}

The corrections to the physical observables of an electron system due to the superconducting fluctuations are the subject of research with more than 50 years old history (see Refs. \cite{LVbook,VarlamovRMP} for a review). Recently the study of superconducting fluctuations has gained a significance as a tool to elucidate the fundamental aspects of
a superconducting state. The conductivity in the normal state is among physical observables which are affected significantly by superconducting fluctuations. 
Near the superconducting transition temperature $T_c$, the most substantial contributions to the dc conductivity  are due to Aslamazov--Larkin \cite{Schmidt,AL1968} and Maki--Thompson \cite{M1968,T1970} processes. While the dc conductivity is sensitive to the position of $T_c$ only, the ac conductivity contains information about the energy and time scales involved. The experimental studies of the microwave conductivity near the superconducting transition in thin films were pioneered in Refs. \cite{dAiello,Lehoczky,Lehoczky2}. Recently, the ac conductivity measurements have been used to elucidate physics behind superconductor-insulator transitions in thin films \cite{Ohashi2006,Armitage2011,Baturina2012,Mondal2013,Frydman2014,Scheffler2016}.

It is well understood theoretically \cite{AVR1983} that the conductivity corrections due to superconducting fluctuations in disordered electron systems in the diffusive regime stand in the same row as the weak localization \cite{GLK1980} and Altshuler--Aronov corrections \cite{AA1979}. The contributions to the conductivity due to pairing fluctuations are nothing but quantum corrections due to interaction in the Cooper channel dressed by the scattering off a random potential. Although the final expressions for the weak localization and Altshuler--Aronov corrections to the ac conductivity $\sigma(\omega)$ are well established \cite{LeeReview,AAbook}, the corresponding expression for the contribution due to superconducting fluctuations is still absent in the literature. The point is that in the diagrammatic approach the pairing conductivity is given by the sum of ten diagrams \cite{LVbook,VarlamovRMP}. Some diagrams produce contributions proportional to $1/(i\omega)$ in the static limit, $\omega \to 0$. However, the sum of all ten diagrams is expected to give a finite contribution to the dc conductivity in the normal state. 
Only recently this problem has been finally solved and the general expression for the superconducting pairing contribution to the dc conductivity  has been established. For the diffusive regime it was derived with the help of the Keldysh path integral and Usadel equation \cite{KSF2012}. In the ballistic regime the fluctuation corrections to the dc conductivity were computed by means of a standard diagrammatic approach \cite{StepanovSkvortsov}. An attempt to obtain a general expression for the fluctuation correction to the ac conductivity, $\sigma(\omega)$, was performed in Ref. \cite{PV} with the help of the Keldysh nonlinear sigma model (see Ref. \cite{KamenevLevchenko} for a review). 
However, the expression derived in Ref. \cite{PV} diverges as $1/(i\omega)$ in the static limit, $\omega\to 0$.

In this paper we report the general analytical expression for the quantum correction to the ac conductivity of a disordered electron system in the diffusive regime which includes the weak localization and Altshuler--Aronov contributions and contributions due to superconducting fluctuations above the transition temperature. We derived our results with the help of the replica Finkel'stein nonlinear sigma model (NL$\sigma$M) (see Refs. \cite{Fin,KB} for a review). In order to find $\sigma(\omega)$ we performed the analytic continuation from Matsubara to real frequencies. We emphasize that our result for the contributions to $\sigma(\omega)$ due to superconducting fluctuations has no $1/(i\omega)$ divergence as $\omega\to 0$. 
In the static limit, $\omega\to 0$, our expression reproduces the results reported for the dc conductivity in Refs. \cite{KSF2012,StepanovSkvortsov}.

The outline of the paper is as follows. In Sec. \ref{s2} we introduce the formalism of the Finkel'stein NL$\sigma$M. The results of the one-loop computation of the ac conductivity are given in Sec. \ref{s3}. In Sec. \ref{s4} the behavior of different contributions to the ac conductivity due to superconducting fluctuations is analysed. 
We finish the paper with conclusion (Sec. \ref{s5}). Some technical details are given in Appendices.

\section{Formalism}
\label{s2}

\subsection{Finkel'stein NL$\sigma$M action}

The action of the Finkel'stein NL$\sigma$M is given as the sum of the non-interacting NL$\sigma$M, $S_\sigma$, 
and contributions due to electron-electron interactions, $S_{\rm int}^{(\rho)}$ (the particle-hole singlet channel), $S_{\rm int}^{(\sigma)}$ (the particle-hole triplet channel), and $S_{\rm int}^{(c)}$ (the particle-particle channel) (see Refs. \cite{Fin,KB,Burmistrov2019} for a review):
\begin{gather}
S=S_\sigma + S_{\rm int}^{(\rho)}+S_{\rm int}^{(\sigma)}+S_{\rm int}^{(c)} ,
\label{eq:NLSM}
\end{gather}
where
\begin{subequations}
\begin{align}
S_\sigma & = -\frac{g}{32} \int d\bm{r} \Tr (\nabla Q)^2 + 4\pi T Z_\omega \int d\bm{r} \Tr \eta Q ,
\label{Ss} \\
S_{\rm int}^{(\rho)} & =-\frac{\pi T}{4} \Gamma_s \!\sum_{\alpha,n} \!\sum_{r=0,3}\!
\int \!\!d\bm{r} \Tr I_n^\alpha t_{r0} Q \Tr I_{-n}^\alpha t_{r0} Q,
\label{Srho}\\
S_{\rm int}^{(\sigma)}& =-\frac{\pi T}{4} \Gamma_t \!\sum_{\alpha,n}\! \sum_{r=0,3}
\int \!\!d\bm{r} \Tr I_n^\alpha \bm{t_r} Q \Tr I_{-n}^\alpha \bm{t_r} Q ,
\label{Ssigma}\\
S_{\rm int}^{(c)}& = -\frac{\pi T}{4}  \Gamma_c\! \sum_{\alpha,n}\! \sum_{r=1,2}  \int \!\!d\bm{r} \Tr  t_{r0} L_n^\alpha Q \Tr t_{r0} L_n^\alpha Q  . 
\label{Sc}
\end{align}
\end{subequations}
Here the matrix field $Q(\bm{r})$ (as well as the trace $\Tr$) acts in the replica, Matsubara, spin, and particle-hole spaces. The matrix field obeys the following constraints:
\begin{gather}
Q^2=1, \quad \Tr Q = 0, \qquad Q^\dag = C^T Q^T C ,
\end{gather}
where the charge-conjugation is realized by the matrix $C=i t_{12}$. 
The action of the NL$\sigma$M involves four constant matrices:
\begin{align}
\Lambda_{nm}^{\alpha\beta} & = \sgn n \, \delta_{nm} \delta^{\alpha\beta}t_{00}, \,
(I_k^\gamma)_{nm}^{\alpha\beta} =\delta_{n-m,k}\delta^{\alpha\beta}\delta^{\alpha\gamma} t_{00}, \notag \\
(L_k^\gamma)_{nm}^{\alpha\beta}& =\delta_{n+m,k}\delta^{\alpha\beta}\delta^{\alpha\gamma} t_{00},
\quad \eta_{nm}^{\alpha\beta} =n \, \delta_{nm}\delta^{\alpha\beta} t_{00} ,
\end{align}
where $\alpha,\beta = 1,\dots, N_r$ stand for replica indices and integers $n,m$ correspond to the
Matsubara fermionic frequencies $\varepsilon_n = \pi T (2n+1)$. The sixteen matrices,
\begin{equation}
\label{trj}
t_{rj} = \tau_r\otimes s_j, \qquad r,j = 0,1,2,3  ,
\end{equation}
operate in the particle-hole (subscript $r$) and spin (subscript $j$) spaces. The matrices 
$\tau_0, \tau_1, \tau_2, \tau_3$ and $s_0, s_1, s_2, s_3$ are the standard sets of the Pauli matrices. Also we introduced the vector $\bm{t_r} =\{t_{r1},t_{r2},t_{r3}\}$ for convenience.

The bare value of the total conductivity 
(in units $e^2/h$ and including spin) is denoted as $g$. 
The interaction amplitude $\Gamma_s$ ($\Gamma_t$) encodes interaction in the singlet (triplet) particle-hole channel. The interaction in the Cooper channel is expressed by $\Gamma_c$. Its negative magnitude, $\Gamma_c<0$, corresponds to an attraction in the particle-particle channel. The parameter $Z_\omega$ describes the frequency renormalization. If Coulomb interaction is present the following relation holds, $\Gamma_s=-Z_\omega$. This condition remains intact under action of the renormalization group flow \cite{Fin,Burmistrov2015}.

\subsection{Kubo formula for the ac conductivity}

Within the Finkel'stein NL$\sigma$M approach, the physical observables, associated with the mean-field parameters of the action \eqref{eq:NLSM}, can be written as correlation functions of the matrix field $Q$. The ac conductivity $\sigma(\omega)$ can be obtained after the analytic continuation to the real frequencies, $i\omega_n \to \omega+i0^+$, of the following Matsubara response function ($\omega_n=2\pi Tn$):
\begin{align}
\sigma(i\omega_n) = & -\frac{g}{16 n} \Bigl \langle \Tr [J_n^\alpha,Q(\bm{r})] [J_{-n}^\alpha,Q(\bm{r})] \Bigr \rangle 
+\frac{g^2}{64 d n} \int d\bm{r}^\prime \Bigl \langle \Tr J_n^\alpha Q(\bm{r}) \nabla Q(\bm{r})
\Tr J_{-n}^\alpha Q(\bm{r}^\prime) \nabla Q(\bm{r}^\prime) \Bigr \rangle  .
\label{eq:PO:g}
\end{align}
Here the expectation values $\langle \dots \rangle$ are taken with respect to the action \eqref{eq:NLSM}, $d$ stands for the spatial dimensionality, and the matrix $J_n^\alpha$ is defined as follows
\begin{equation}
J_n^\alpha = \frac{t_{30}-t_{00}}{2} I_n^\alpha + \frac{t_{30}+t_{00}}{2} I_{-n}^\alpha .
\end{equation}
At the classical level, $Q=\Lambda$, the conductivity is independent of the frequency, $\sigma(\omega) = g$.

\section{One-loop corrections to the ac conductivity}
\label{s3}

\subsection{Perturbative expansion}

Our aim is to compute correction to $\sigma(\omega)$ in the lowest order in $1/g$. For this purpose we shall use the square-root parametrization of the matrix field $Q$:
\begin{gather}
Q = W +\Lambda \sqrt{1-W^2}, \qquad W= \begin{pmatrix}
0 & w\\
\overline{w} & 0
\end{pmatrix} .
\label{eq:Q-W}
\end{gather}
We adopt the following notations: $W_{n_1n_2} = w_{n_1n_2}$ and $W_{n_2n_1} = \overline{w}_{n_2n_1}$ where $n_1\geqslant 0$ and $n_2< 0$.
The blocks $w$ and $\overline{w}$ satisfy the charge-conjugation constraints:
\begin{gather}
\overline{w} = -C w^T C,\qquad w = - C w^* C .
\end{gather}
These constraints imply that some elements $(w^{\alpha\beta}_{n_1n_2})_{rj}$ in the expansion, $w^{\alpha\beta}_{n_1n_2}= \sum_{rj} (w^{\alpha\beta}_{n_1n_2})_{rj} t_{rj}$, are purely real and the others are purely imaginary.

The part of the action \eqref{eq:NLSM}, which is  quadratic in $W$, determines the following propagators for diffusive modes in the theory. The propagators of \emph{diffusons} (modes with $r=0,3$ and $j=0,1,2,3$) read
\begin{gather}
\Bigl \langle [w_{rj}(\bm{p})]^{\alpha_1\beta_1}_{n_1n_2} [\bar{w}_{rj}(-\bm{p})]^{\beta_2\alpha_2}_{n_4n_3} \Bigr \rangle =  \frac{2}{g} \delta^{\alpha_1\alpha_2} \delta^{\beta_1\beta_2}\delta_{n_{12},n_{34}}\mathcal{D}_p(i\Omega_{12}^\varepsilon)\Bigl [\delta_{n_1n_3} - \frac{32 \pi T \Gamma_j}{g}\delta^{\alpha_1\beta_1}  \mathcal{D}_p^{(j)}(i\Omega_{12}^\varepsilon) \Bigr ] ,
\label{eq:prop:PH}
\end{gather}
where $\Omega_{12}^\varepsilon = \varepsilon_{n_1}-\varepsilon_{n_2}=2\pi T n_{12}=2\pi T (n_1-n_2)$, $\Gamma_0\equiv \Gamma_s$, and $\Gamma_1=\Gamma_2=\Gamma_3\equiv \Gamma_t$. The diffuson in the absence of interaction is given as
\begin{equation}
\mathcal{D}^{-1}_p(i\omega_n) =p^2+{16 Z_\omega |\omega_n|}/{g} .
\end{equation}
The diffusons renormalized by a ladder resummation of interaction in the singlet  and triplet particle-hole channels have the following form, respectively,
\begin{align}
\mathcal{D}^{(0)}_p(i\omega_n)\equiv \mathcal{D}^s_p(i\omega_n) & =  \Bigl [p^2+{16 (Z_\omega+\Gamma_s) |\omega_n|}/{g}\Bigr ]^{-1},\notag \\
 \mathcal{D}^{(1,2,3)}_p(i\omega_n)\equiv \mathcal{D}^t_p(i\omega_n) & =  \Bigl [p^2+{16 (Z_\omega+\Gamma_t) |\omega_n|}/{g}\Bigr ]^{-1} .
\end{align}
The propagators of \emph{singlet cooperons} (modes with $r=1,2$ and $j=0$) can be written as
\begin{gather}
\Bigl \langle [w_{r0}(\bm{p})]^{\alpha_1\beta_1}_{n_1n_2} [\bar{w}_{r0}(-\bm{p})]^{\beta_2\alpha_2}_{n_4n_3} \Bigr \rangle =  \frac{2}{g} \delta^{\alpha_1\alpha_2} \delta^{\beta_1\beta_2}\delta_{n_{14},n_{32}}
\mathcal{C}_p(i\Omega_{12}^\varepsilon)\Bigl [\delta_{n_1n_3} - \frac{4 \pi T }{D}\delta^{\alpha_1\beta_1}  \mathcal{C}_p(i\Omega_{34}^\varepsilon)  \mathcal{L}_p(i\mathcal{E}_{12}) \Bigr ] ,
\label{eq:prop:PPS}
\end{gather}
where $\mathcal{E}_{12} = \varepsilon_{n_1}+\varepsilon_{n_2}$, $\mathcal{C}_p(i\omega_n) \equiv \mathcal{D}_p(i\omega_n)$. The diffusion coefficient is $D = g/(16 Z_\omega)$. The fluctuation propagator has the standard form,
\begin{gather}
\mathcal{L}^{-1}_p(i\omega_n) = \gamma_c^{-1}
- \ln (2\pi T\tau) - \psi\left (\mathcal{X}_{p,i|\omega_n|} \right )
 + \psi\left ({1}/{2} \right ) ,
 \label{eq:def:fl:prop}
\end{gather}
where $\gamma_c= \Gamma_c/Z_\omega$ and $\psi(z)$ denotes the di-gamma function. Also we introduced the following notation
\begin{equation}
\mathcal{X}_{q,\omega} = \frac{D q^2- i\omega}{4\pi T}+\frac{1}{2}  .
\end{equation}
The \emph{triplet cooperons} (modes with $r=1,2$ and $j=1,2,3$) are insensitive to the Cooper-channel interaction and coincide with the non-interacting cooperons:
\begin{align}
\Bigl \langle [w_{rj}(\bm{p})]^{\alpha_1\beta_1}_{n_1n_2} [\bar{w}_{rj}(-\bm{p})]^{\beta_2\alpha_2}_{n_4n_3} \Bigr \rangle & =  \frac{2}{g} \delta^{\alpha_1\alpha_2} \delta^{\beta_1\beta_2} \delta_{n_1n_3}
 \delta_{n_2n_4}\mathcal{C}_p(i\Omega_{12}^\varepsilon) .
 \label{eq:prop:PPT}
\end{align}

\subsection{One-loop renormalization}

Expanding the matrix $Q$ up to the second order in $W$ we obtain the following expression from Eq. \eqref{eq:PO:g},
\begin{gather}
\sigma(i\omega_n) =  g  -\frac{g}{64 n} \Bigl \langle \Tr [J_n^\alpha,\Lambda W^2(\bm{r})] [J_{-n}^\alpha,\Lambda W^2(\bm{r})] \Bigr \rangle  +\frac{g^2}{64 d n} \int d\bm{r}^\prime \Bigl \langle \Tr J_n^\alpha W(\bm{r}) \nabla W(\bm{r})
\notag \\
\times
\Tr J_{-n}^\alpha W(\bm{r}^\prime) \nabla W(\bm{r}^\prime) \Bigr \rangle  .
\label{app:eq:PO:g}
\end{gather}
In order to derive the correction to $\sigma(i\omega_n)$ in the lowest order in $1/g$, it is enough to average the correlation functions in Eq. \eqref{app:eq:PO:g} with the Gaussian part of the NL$\sigma$M action. Using Wick theorem and computing the averages with the help of Eqs. \eqref{eq:prop:PH} - \eqref{eq:prop:PPT}, we find lengthy expression
\begin{subequations}
\begin{align}
\sigma(i\omega_n)  = & g - 4 \int_q \mathcal{C}_q(i\omega_n)
 -  \frac{16 \pi^2 T^2}{\omega_n D} \sum_{\omega_n>\varepsilon_{n_1},-\varepsilon_{n_2}>0}
\int_q \mathcal{C}_q(i\Omega_{12}^\varepsilon)
\mathcal{C}_q(2i\omega_n-i\Omega_{12}^\varepsilon)\mathcal{L}_q(i\mathcal{E}_{12})
\label{app:cond:WL+MTan} \\
+ & \frac{256 \pi T}{\omega_n g d} \sum_{j=0}^3 \Gamma_j \sum_{\omega_m>0} \int_q \, q^2
\min\{\omega_m,\omega_n\} \mathcal{D}_q(i\omega_m) \mathcal{D}^{(j)}_q(i\omega_m)\mathcal{D}_q(i\omega_m+i\omega_n)
\label{app:cond:AA} \\
 - & \frac{16 \pi^2 T^2}{\omega_n D} \sum_{\varepsilon_{n_1},-\varepsilon_{n_2}>0}
\sum_{\sigma,\sigma^\prime=\pm} \int_q  \mathcal{C}_q\left (i\Omega_{12}^\varepsilon+i\omega_n\zeta^2_{\sigma\sigma^\prime}\right )\mathcal{C}_q\left (i\Omega_{12}^\varepsilon+i\omega_n(2-\zeta^2_{\sigma\sigma^\prime})\right )
\notag\\
& \,{}\hspace{3.5cm} \times
\mathcal{L}_q\left (i\mathcal{E}_{12}+i\omega_n \zeta_{\sigma\sigma^\prime}\right )
\label{app:cond:MTreg} \\
 + & \frac{32 \pi^2 T^2}{d \omega_n D} \sum_{\varepsilon_{n_1},-\varepsilon_{n_2}>0}
\sum_{\sigma,\sigma^\prime=\pm} \int_q  q^2 \mathcal{C}_q\left (i\Omega_{12}^\varepsilon\right )\mathcal{C}_q\left (i\Omega_{12}^\varepsilon+i\omega_n\right )\mathcal{C}_q\left (i\Omega_{12}^\varepsilon+i\omega_n(2-\zeta^2_{\sigma\sigma^\prime})\right )
\notag \\
& \,{} \hspace{3.5cm} \times
\mathcal{L}_q\left (i\mathcal{E}_{12}+i\omega_n \zeta_{\sigma\sigma^\prime}\right )
\label{app:cond:Dos1} \\
 + & \frac{32 \pi^2 T^2}{d \omega_n D} \sum_{\varepsilon_{n_1},-\varepsilon_{n_2}>0}
\sum_{\sigma,\sigma^\prime=\pm} \int_q  q^2 \mathcal{C}_q\left (i\Omega_{12}^\varepsilon\right )\mathcal{C}_q\left (i\Omega_{12}^\varepsilon+i\omega_n\right )\mathcal{C}_q\left (i\Omega_{12}^\varepsilon+i\omega_n(1+\sigma)\right )
\notag \\
& \,{} \hspace{3.5cm} \times
\mathcal{L}_q\left (i\mathcal{E}_{12}\right )
\label{app:cond:Dos2} \\
 - & \frac{128 \pi^3 T^3 z}{d \omega_n D^2} \sum_{\varepsilon_{n_{1,3}},-\varepsilon_{n_{2,4}}>0}
 \sum_{\sigma,\sigma^\prime=\pm} \int_q  q^2 \mathcal{C}_q\left (i\Omega_{12}^\varepsilon\right )\mathcal{C}_q\left (i\Omega_{12}^\varepsilon+i\omega_n\right )
\mathcal{C}_q\left (i\Omega_{34}^\varepsilon\right )\mathcal{C}_q\left (i\Omega_{34}^\varepsilon+i\omega_n\right )
\notag \\
& \,{} \hspace{3.5cm} \times\delta_{\mathcal{E}_{12},\mathcal{E}_{34}+i\omega_n \mu^-_{\sigma\sigma^\prime}}
\mathcal{L}_q\left (i\mathcal{E}_{12}\right )
\mathcal{L}_q\left (i\mathcal{E}_{34}+i\omega_n \mu^+_{\sigma\sigma^\prime}\right ) .
\label{app:cond:AL}
\end{align}
\end{subequations}
Here we use the following short-hand notations, $\zeta_{\sigma\sigma^\prime} = (\sigma+\sigma^\prime)/2$, $\mu_{\sigma\sigma^\prime}^{\pm} =
\sigma(1\pm \sigma^\prime)/2$, and 
$ \int_q \equiv \int {d^d\bm{q}}/{(2\pi)^d}$.
We note that the contributions 
\eqref{app:cond:WL+MTan} and \eqref{app:cond:MTreg} come from the term  in Eq. \eqref{app:eq:PO:g} which has no gradients acting on $W$ matrices. All the other contributions result from the last term in the right hand side of Eq. \eqref{app:eq:PO:g}.   

Traditionally, the conductivity is split into several parts: weak localization or interference contribution $\delta g^{WL}$, Altshuler--Aronov or interaction contribution $\delta g^{AA}$, and fluctuation conductivity which stems from the interaction in the Cooper channel, $\delta g^{CC}$, i.e.
\begin{equation}
\sigma(\omega)  = g + \delta g^{\rm WL}(\omega)+\delta g^{\rm AA}(\omega)+\delta g^{\rm CC}(\omega) .
\end{equation}
The contribution due to the Cooper channel interaction involves  the fluctuation propagator $\mathcal{L}_q$. This contribution, $\delta g^{\rm CC}$, can be written as a sum of four terms~\cite{LVbook}:
\begin{equation}
\delta g^{\rm CC} = \delta g^{\rm MT, an} +  \delta \tilde{g}^{\rm MT, reg} + \delta \tilde{g}^{\rm DOS} + \delta\tilde{g}^{\rm AL} .
\end{equation}
In what follows we shall consider each of these terms separately.

\subsubsection{Weak localization and Althsuler--Aronov corrections}

The weak localization and Althsuler--Aronov contributions are given by the second term in the right hand side of Eq.\eqref{app:cond:WL+MTan} and by Eq. \eqref{app:cond:AA}. At first, we perform analytic continuation to the real frequencies $i\omega_n\to \omega+i0^+$. Then the interference correction is expressed in terms of the non-interacting cooperon \cite{GLK1980}:
\begin{equation}
\delta g^{\rm WL}(\omega)=- 4 \int_q \mathcal{C}^R_q(\omega) .
\label{app:WL}
\end{equation}
Here $\mathcal{C}^R_q(\omega)$ stands for the retarded propagator corresponding to the Matsubara propagator $\mathcal{C}_q(i\omega_n)$.

The interaction correction reads \cite{AA1979,Fin198384a,Fin198384b,Fin198384c,CCLM1984}
\begin{gather}
\delta g^{\rm AA}(\omega)= \frac{64}{i \omega g d} \sum_{j=0}^3 \Gamma_j \int_{q,\Omega} \, q^2
\Bigl [ \Omega \mathcal{B}_\Omega - (\Omega-\omega) \mathcal{B}_{\Omega-\omega}\Bigr ]
 \mathcal{D}^R_q(\Omega) \mathcal{D}^{(j),R}_q(\Omega)\mathcal{D}^R_q(\Omega+\omega).
 \label{app:AA}
\end{gather}
Here $\mathcal{B}_\Omega=\coth[\Omega/(2T)]$ denotes the bosonic distribution function for the particle-hole excitations. The retarded diffuson propagators are denoted as $\mathcal{D}^R_q(\omega)$, $\mathcal{D}^{(j),R}_q(\omega)$. Also we introduced the short-hand notation $\int_\Omega\equiv \int_{-\infty}^\infty d\Omega $.

\subsubsection{Anomalous Maki-Thompson correction}

The anomalous Maki-Thompson correction \cite{M1968,T1970} is given by the last term in the right hand side of Eq. \eqref{app:cond:WL+MTan}. It is convenient to rewrite it as follows
\begin{equation}
\delta g^{\rm MT, an}(i\omega_n) =  4 \int_q \mathcal{C}_q(i\omega_n) \beta_q(i\omega_n) ,
\end{equation}
where~\cite{AV1979,L1980}
\begin{gather}
\beta_q(i\omega_n) =  \frac{\pi T}{\omega_n} \sum_{|\omega_m|<\omega_n}
\mathcal{L}_q(i\omega_m) \Bigl [\psi(\mathcal{X}_{q,i|\omega_m|}) - \psi(\mathcal{X}_{q,2i\omega_n-i|\omega_m|}) \Bigr ] .
\end{gather}

Performing analytic continuation to the real frequencies we obtain the final form of the anomalous Maki-Thompson correction
\begin{equation}
\delta g^{\rm MT, an}(\omega) = 4 \int_q \mathcal{C}_q^R(\omega) \beta_q^R(\omega) ,
\label{app:MT:an:t}
\end{equation}
where
\begin{gather}
\beta_q^R(\omega)  =
\int_{\Omega} \mathcal{L}_q^R(\Omega) \frac{\mathcal{B}_\Omega-\mathcal{B}_{\Omega-\omega} }{2\omega}
 \Bigl [ \psi(\mathcal{X}_{q,\Omega})- \psi(\mathcal{X}_{q,2\omega-\Omega})
\Bigr ]
.
\end{gather}
We mention that the anomalous Maki-Thompson correction \eqref{app:MT:an:t}
 coincides with the sum $\sigma_{MT1}+ \sigma_{MT2}$ computed in Ref. \cite{PV} (see Eqs. (A1) and (A2) there).

It is instructive to compare the above result with the other expressions existing in the literature. For this purpose we use the following relations
\begin{equation}
\int_\varepsilon \Bigl(\mathcal{F}_{\varepsilon+\omega}- \mathcal{F}_\varepsilon\Bigr ) \ \mathcal{C}_q^{R}(2\varepsilon+\Omega)
= iD\bigl[\psi(\mathcal{X}_{q,\Omega})- \psi(\mathcal{X}_{q,\Omega-2\omega})\bigr ]
\label{eq:rel1}
\end{equation}
and
\begin{gather}
\int_\varepsilon \mathcal{F}_{\varepsilon+\Omega} \Bigl(\mathcal{F}_{\varepsilon+\omega}- \mathcal{F}_\varepsilon\Bigr )\  \mathcal{C}_q^{R}(2\varepsilon+\Omega) = iD \Bigl\{\mathcal{B}_\Omega 
\bigl [
\psi(\mathcal{X}_{q,\Omega})
-\psi(\mathcal{X}_{q,-\Omega})
\bigr ]
\notag \\
-\mathcal{B}_{\Omega-\omega}
\bigl[ \psi(\mathcal{X}_{q,\Omega-2\omega})
- \psi(\mathcal{X}_{q,-\Omega})\bigr ]
\Bigr \} .
\label{eq:rel2}
\end{gather}
Here $\mathcal{F}_\varepsilon=\tanh[\varepsilon/(2T)]$ stands for the fermionic distribution function. Then it is possible to rewrite Eq. \eqref{app:MT:an:t} as follows 
\begin{gather}
\delta g^{\rm MT, an}(\omega)  =
\frac{2 i}{D\omega} 
\int_{q,\Omega,\varepsilon} \mathcal{C}_q^R(\omega) \mathcal{L}_q^R(\Omega) \Bigl[\mathcal{F}_{\varepsilon+\omega} - \mathcal{F}_\varepsilon\Bigr]
\Bigl[\mathcal{B}_{\Omega} - \mathcal{F}_{\varepsilon+\Omega}\Bigr]
\mathcal{C}^R_q(2\varepsilon+\Omega) .
\label{app:MT:an:final:diff}
\end{gather}
In the dc limit, $\omega\to 0$, the expression \eqref{app:MT:an:final:diff} is similar to Eq. (384) of Ref. \cite{KamenevLevchenko}.

\subsubsection{Regular Maki-Thompson correction}

The so-called regular part of the Maki-Thompson correction is determined by the contribution
\eqref{app:cond:MTreg}. Performing summation over one of the fermionic energies, we obtain
\begin{gather}
 \delta \tilde{g}^{\rm MT, reg}(i\omega_n)  =
-\frac{D}{\omega_n}\int_q  \sum_{\omega_m}  \Biggl \{ 2\psi^\prime \left (\mathcal{X}_{q,i|\omega_{m+ n}|+i\omega_n}\right )
 +
 \frac{4\pi T}{\omega_n}
\Bigl [
\psi \left (\mathcal{X}_{q,i|\omega_{m}|+2i\omega_n}\right )
 -
\psi \left (\mathcal{X}_{q,i|\omega_{m}|}\right )
\Bigr ]
\Biggr \}  \mathcal{L}_q(i\omega_m)
.
\end{gather}
After the analytic continuation to the real frequency, $i\omega_n\to \omega+i0$, we find 
\begin{gather}
\delta \tilde{g}^{\rm MT, reg}(\omega) =
- \frac{2D}{\pi T\omega}  \int_{q,\Omega} \mathcal{B}_\Omega \mathcal{L}_q^R(\Omega) \psi^{\prime}(\mathcal{X}_{q,\Omega})
+\frac{D}{2\pi T \omega} \int_{q,\Omega} 
\mathcal{L}_q^R(\Omega) \Biggl\{ 2 \mathcal{B}_\Omega
\Phi_{-2\omega}(\Omega)
\notag \\
+ \mathcal{B}_\Omega
\Bigl[ \psi^\prime(\mathcal{X}_{q,\Omega})-\psi^\prime(\mathcal{X}_{q,\Omega+2\omega})
\Bigr]
+  \Bigl[ \mathcal{B}_\Omega-\mathcal{B}_{\Omega-\omega}\Bigr ]
\Bigl[ \psi^\prime(\mathcal{X}_{q,\Omega})
-\psi^\prime(\mathcal{X}_{q,2\omega-\Omega})
\Bigr] \Biggr\} ,
\label{app:MT:reg:t}
\end{gather}
where 
\begin{equation}
\Phi_\omega(\Omega) =
\psi^\prime(\mathcal{X}_{q,\Omega}) + \frac{4\pi T}{i\omega}
\Bigl [ \psi(\mathcal{X}_{q,\Omega})-
\psi(\mathcal{X}_{q,\Omega-\omega})
\Bigr ] .
\end{equation}
We note that in the course of derivation of Eq. \eqref{app:MT:reg:t} we have also used the following symmetry properties: $\mathcal{L}_q^A(\Omega)
= \mathcal{L}_q^R(-\Omega)$, and $\mathcal{B}_{-\Omega}=-\mathcal{B}_{\Omega}$.

It is useful to relate the regular Maki-Thompson correction with the correction to the tunneling density of states due to interaction in the Cooper channel \cite{Abrahams1970,Castro1990}. The correction to the density of states can be written as~\cite{KamenevLevchenko}
\begin{equation}
\delta\rho^{\rm CC}(\varepsilon) = 
\rho_0 \re \Upsilon(\varepsilon) ,
\end{equation}
where
\begin{equation}
\Upsilon(\varepsilon)  = \frac{32 Z_\omega}{i g^2}  \int_{q,\Omega}  \mathcal{C}^{R2}_q(2\varepsilon - \Omega)  \Bigl [\mathcal{L}^K_q(\Omega)+\mathcal{F}_{\varepsilon-\Omega} \mathcal{L}^R_q(\Omega) \Bigr ] .
\end{equation}
Here $\mathcal{L}^K_q(\Omega)=2i \mathcal{B}_{\Omega}\im \mathcal{L}^R_q(\Omega)$ stands for the Keldysh component of the fluctuation propagator. 

We define the correction to the conductivity that is related with the correction to the density of states in the following way
\begin{equation}
\delta g^{\rm DOS}(\omega) = \frac{g}{\omega} \int d\varepsilon \Bigl [ f_F(\varepsilon-\omega)-f_F(\varepsilon)\Bigr ] \Upsilon(\varepsilon) ,
\label{eq_Sigma_DOS-PP}
\end{equation}
where $f_F(\varepsilon)=(1-\mathcal{F}_\varepsilon)/2$ is the Fermi-Dirac distribution function. Then, using the identities \eqref{eq:rel1} and \eqref{eq:rel2},
we obtain the following result
\begin{gather}
\delta g^{\rm DOS}(\omega) = \frac{D}{4\pi T\omega}  \int_{q,\Omega} \mathcal{L}^R_q(\Omega)
\Biggl\{ \Bigl[ \mathcal{B}_\Omega-\mathcal{B}_{\Omega-\omega}
 \Bigr ]
\Bigl [ \psi^{\prime}\left (\mathcal{X}_{q,\Omega}\right )
 - \psi^{\prime}\left (\mathcal{X}_{q,-\Omega+2\omega}\right )\Bigr ]
 \notag \\
+\mathcal{B}_\Omega \Bigl [
 \psi^{\prime}\left (\mathcal{X}_{q,\Omega}\right )-
\psi^{\prime}\left (\mathcal{X}_{q,\Omega+2\omega}\right )
\Bigr ]
\Biggr \} .
\label{eq_Sigma_DOS}
\end{gather}
Next, using Eq. \eqref{eq_Sigma_DOS}, we split the regular Maki-Thompson contribution into three parts
\begin{gather}
\delta \tilde{g}^{\rm MT, reg}(\omega) =
- \frac{2D}{\pi T\omega}  \int_{q,\Omega} \mathcal{B}_\Omega \mathcal{L}_q^R(\Omega) \psi^{\prime}(\mathcal{X}_{q,\Omega})
+ \delta {g}^{\rm DOS}(\omega) +
\delta {g}^{\rm sc, 1}(\omega) ,
\label{app:MT:reg:t:final}
\end{gather}
where
\begin{gather}
\delta {g}^{\rm sc, 1}(\omega)
=\frac{D}{4\pi T \omega} \int_{q,\Omega} 
\mathcal{L}_q^R(\Omega) \Biggl\{ 4 \mathcal{B}_\Omega \Phi_{-2\omega}(\Omega)
+ \mathcal{B}_\Omega
\Bigl[ \psi^\prime(\mathcal{X}_{q,\Omega})-\psi^\prime(\mathcal{X}_{q,\Omega+2\omega})
\Bigr]
\notag\\
+  \Bigl[ \mathcal{B}_\Omega-\mathcal{B}_{\Omega-\omega}\Bigr ]
\Bigl[ \psi^\prime(\mathcal{X}_{q,\Omega})
-\psi^\prime(\mathcal{X}_{q,2\omega-\Omega})
\Bigr] \Biggr\}.
\label{eq:sigma:sc1}
\end{gather}

\subsubsection{DOS-type correction}

The so-called DOS-type correction~\cite{LVbook} is given by contributions \eqref{app:cond:Dos1} - \eqref{app:cond:Dos2}.
It is convenient to rewrite them as follows
\begin{gather}
\delta \tilde{g}^{\rm DOS}(i\omega_n) = \frac{4D^2}{\omega_n^2 d}
\int_q q^2 \sum_{\omega_m} \mathcal{L}_q(i\omega_m)
\Biggl \{
\psi^\prime\left (\mathcal{X}_{q,i|\omega_{m}|} \right )
-
\psi^\prime\left (\mathcal{X}_{q,i|\omega_{m+n}|+i\omega_n} \right )
+\frac{4\pi T}{\omega_n}
\Bigl [
\psi\left (\mathcal{X}_{q,i|\omega_{m+n}|+i\omega_n} \right )
\notag \\
-
\psi\left (\mathcal{X}_{q,i|\omega_{m+n}|}\right )
+\psi\left (\mathcal{X}_{q,i|\omega_{m}|+i\omega_n}\right )
- \psi\left (\mathcal{X}_{q,i|\omega_{m}|+2i\omega_n}\right )
\Bigr ]
\Biggr \} .
\label{app:eq:11}
\end{gather}
The analytic continuation of Eq. \eqref{app:eq:11} to the real frequency, $i\omega_n\to \omega+i0$, yields
\begin{gather}
\delta \tilde{g}^{\rm DOS}(\omega)  =\frac{i D^2}{\pi T\omega^2 d} 
\int_{q,\Omega}  q^2 \mathcal{L}_q^R(\Omega)
\Biggl \{
\bigl (\mathcal{B}_\Omega-\mathcal{B}_{\Omega-\omega} \bigr )  \Bigl[ \Phi_\omega(\Omega) - \Phi_\omega(2\omega-\Omega)\Bigr ]
+
2 \mathcal{B}_\Omega \Bigl [
\psi^\prime(\mathcal{X}_{q,\Omega})
\notag \\
-
\psi^\prime(\mathcal{X}_{q,\Omega+2\omega})
\Bigr ]
+ \mathcal{B}_\Omega
\Bigl[ \Phi_\omega(2\omega+\Omega) -\Phi_\omega(\Omega) \Bigr ]\Biggr \} .
\end{gather}
It is useful to single out explicitly the part that diverges in the limit $\omega\to 0$. Then we obtain
\begin{gather}
\delta \tilde{g}^{\rm DOS}(\omega) =-\frac{D^2}{d\pi^2 T^2 \omega}
\int_{q,\Omega} q^2  \mathcal{B}_\Omega \mathcal{L}_q^R(\Omega) \psi^{\prime\prime}(\mathcal{X}_{q,\Omega}) 
+ \delta {g}^{\rm sc, 2}(\omega) ,
\label{app:DOS:t:final}
\end{gather}
where
\begin{gather}
\delta {g}^{\rm sc, 2}(\omega)  =\frac{i D^2}{\pi T\omega^2 d} 
\int_{q,\Omega}  q^2 \mathcal{L}_q^R(\Omega)
\Biggl \{
\bigl (\mathcal{B}_\Omega-\mathcal{B}_{\Omega-\omega} \bigr )  \Bigl[ \Phi_\omega(\Omega) - \Phi_\omega(2\omega-\Omega)\Bigr ]
+
2 \mathcal{B}_\Omega \Bigl [
\psi^\prime(\mathcal{X}_{q,\Omega})
\notag \\
-
\psi^\prime(\mathcal{X}_{q,\Omega+2\omega})
- \frac{i\omega}{2\pi T} \psi^{\prime\prime}(\mathcal{X}_{q,\Omega})
\Bigr ]
+ \mathcal{B}_\Omega
\Bigl[ \Phi_\omega(2\omega+\Omega) -\Phi_\omega(\Omega) \Bigr ]\Biggr \} .
\label{eq:sigma:sc2}
\end{gather}

\subsubsection{Aslamazov-Larkin correction}

The contribution
\eqref{app:cond:AL} is the correction due to Aslamazov--Larkin process~\cite{AL1968}. It can be written as follows
\begin{align}
\delta\tilde{g}^{\rm AL}(i\omega_n) & =  - \frac{8\pi T}{d \omega_n} \left ( \frac{D}{4\pi T}\right )^2 \int_q q^2
\sum_{\omega_m}  \mathcal{L}_q(i\omega_m)
\mathcal{L}_q(i\omega_{m+n}) \Delta_q^2(i\omega_m,i\omega_{m+n},i\omega_n) ,
\end{align}
where
\begin{gather}
\Delta_q(i\omega_m,i\omega_{k},i\omega_n)  =   - \frac{4\pi T}{\omega_n}
\Biggl [
\psi \left (\mathcal{X}_{q,i|\omega_{m}|}\right )+ \psi \left (\mathcal{X}_{q,i|\omega_{k}|}\right ) 
 -
\psi \left (\mathcal{X}_{q,i|\omega_{m}|+i\omega_n}\right )
   -
\psi \left (\mathcal{X}_{q,i|\omega_{k}|+i\omega_n}\right )
\Biggr ] .
\end{gather}
After the analytic continuation to the real frequency, $i\omega_n\to \omega+i0$, we find
\begin{gather}
\delta\tilde{g}^{\rm AL}(\omega) =  - \frac{D^2}{8d\pi^2 T^2 \omega} \int_{q,\Omega} q ^2  \mathcal{L}_q^R(\Omega) \Biggl \{
\Bigl ( \mathcal{B}_\Omega+\mathcal{B}_{\Omega-\omega}\Bigr ) 
\mathcal{L}_q^R(\Omega-\omega) \bigl ( \Delta^{RRR}_q(\Omega-\omega,\Omega,\omega)\bigr )^2
\notag \\
+\Bigl (\mathcal{B}_{\Omega-\omega} - \mathcal{B}_\Omega \Bigr )
\Bigl [\mathcal{L}_q^R(\Omega-\omega)\bigl ( \Delta^{RRR}_q(\Omega-\omega,\Omega,\omega)\bigr )^2
 - \mathcal{L}_q^A(\Omega-\omega)\bigl ( \Delta^{ARR}_q(\Omega-\omega,\Omega,\omega)\bigr )^2
\Bigr ]
\Biggr \} .
\label{app:eq:12}
\end{gather}
Here we introduced the function
\begin{align}
\Delta^{RRR}_q(\Omega,\Omega^\prime,\omega) & =
 \frac{4\pi T}{i\omega}
\Biggl [
\psi \left (\mathcal{X}_{q,\Omega}\right )
 - \psi \left (\mathcal{X}_{q,\Omega+\omega}\right )
+ \psi \left (\mathcal{X}_{q,\Omega^\prime}\right )
 - \psi \left (\mathcal{X}_{q,\Omega^\prime+\omega}\right )
\Biggr ] .
\end{align}
The function $\Delta^{ARR}_q(\Omega,\Omega^\prime,\omega)$ can be obtained from $\Delta^{RRR}_q(\Omega,\Omega^\prime,\omega)$ according to the following prescription, $\Delta^{ARR}_q(\Omega,\Omega^\prime,\omega)  = \Delta^{RRR}_q(-\Omega,\Omega^\prime,\omega)$.
We note that Eq. \eqref{app:eq:12} coincides with the general result for the Aslamazov--Larkin contribution computed by the diagrammatic technique (see Eq. (7.105) in Ref. \cite{LVbook}). It is convenient to rewrite the correction \eqref{app:eq:12} in the following way 
\begin{gather}
\delta\tilde{g}^{\rm AL}(\omega) = 
- \frac{D^2}{d\pi^2 T^2 \omega} \int_{q,\Omega} q ^2 
\mathcal{B}_\Omega \Bigl [ \mathcal{L}_q^R(\Omega) 
\psi^\prime \left (X_{q,\Omega}\right )\Bigr ]^2
+
 \delta{g}^{\rm AL}(\omega) 
+ \delta{g}^{\rm sc, 3}(\omega) .
\label{app:AL:t0}
\end{gather}
Here we single out the term which diverges in the limit of zero frequency, $\omega\to 0$.
Next, we introduce
\begin{gather}
\delta{g}^{\rm AL}(\omega) =  - \frac{4D^2}{d \omega^3} \int_{q,\Omega} q ^2  \Bigl (\mathcal{B}_{\Omega-\omega} -
\mathcal{B}_\Omega \Bigr )
\mathcal{L}_q^R(\Omega) \Bigl [ \psi\bigl (\mathcal{X}_{q,\Omega-\omega}\bigr )-\psi\bigl (\mathcal{X}_{q,\Omega+\omega}\bigr )\Bigr] \im \mathcal{L}_q^R(\Omega-\omega)
\notag \\
\times 
\im \Bigl [ \psi\bigl (\mathcal{X}_{q,\Omega-\omega}\bigr )-\psi\bigl (\mathcal{X}_{q,\Omega+\omega}\bigr )\Bigr]
\label{app:AL:t}
\end{gather}
and
\begin{gather}
\delta{g}^{\rm sc, 3}(\omega) =  - \frac{D^2}{8d\pi^2 T^2 \omega} \int_{q,\Omega} q ^2  \mathcal{L}_q^R(\Omega) \Biggl \{
\Bigl (\mathcal{B}_\Omega+\mathcal{B}_{\Omega-\omega}\Bigr ) 
\mathcal{L}_q^R(\Omega-\omega) 
\bigl ( \Delta^{RRR}_q(\Omega-\omega,\Omega,\omega)\bigr )^2- 8 \mathcal{B}_\Omega 
\notag \\
\times  
\mathcal{L}_q^{R}(\Omega) \psi^{\prime 2} \left (X_{q,\Omega}\right )
+\Bigl (\mathcal{B}_{\Omega-\omega} - \mathcal{B}_\Omega \Bigr )
\Biggl [\mathcal{L}_q^A(\Omega-\omega)
\Bigl[  \Delta^{RRR}_q(\Omega-\omega,\Omega,\omega)
\re  \Delta^{RRR}_q(\Omega-\omega,\Omega,\omega) 
\notag \\-
\bigl ( \Delta^{ARR}_q(\Omega-\omega,\Omega,\omega)\bigr )^2
\Bigr ]
 + i \mathcal{L}_q^R(\Omega-\omega)
 \Delta^{RRR}_q(\Omega-\omega,\Omega,\omega)
\im  \Delta^{RRR}_q(\Omega-\omega,\Omega,\omega)
\Biggr ]
\Biggr \} .
\label{app:eq:sc3}
\end{gather}

\subsection{Final result}

Naturally, one expects that the dc conductivity in the normal state of a disordered electron system is finite. We note that separate contributions to $\delta g^{CC}(\omega)$ do not satisfy this requirement. In particular, there are terms in Eqs. \eqref{app:MT:reg:t:final}, \eqref{app:DOS:t:final}, and \eqref{app:AL:t} which diverge as $1/(i\omega)$ in the limit $\omega\to 0$. They can be summed up as follows:
\begin{gather}
- \frac{2D}{\pi T\omega}  \int_{q,\Omega} \mathcal{B}_\Omega \mathcal{L}_q^R(\Omega) \Bigl\{ \psi^{\prime}(\mathcal{X}_{q,\Omega})+ \frac{D q^2}{2 d\pi T} \psi^{\prime\prime}(\mathcal{X}_{q,\Omega})
+  \frac{D q^2}{2 d\pi T}\mathcal{L}_q^R(\Omega) 
\bigl[ \psi^\prime \left (X_{q,\Omega}\right )\bigr ]^2 \Bigr \}\notag 
\\
=\frac{4}{i \omega d}\im  \int_{q,\Omega} 
\mathcal{B}_\Omega \partial_{q_\mu}  \partial_{q_\mu}\ln \mathcal{L}^R_q(\Omega) .
\label{eq:divergent1}
\end{gather}
Thus the sum of all  terms in  $\delta g^{CC}(\omega)$ which are proportional to $1/(i\omega)$ has the form of the total second derivative with respect to the momentum. This implies that the contribution \eqref{eq:divergent1} is determined by the ultraviolet and, consequently, cannot be accurately computed within NL$\sigma$M that is the low-energy effective theory only. However, as one can check \cite{StepanovSkvortsov}, the contribution from the ballistic scales has exactly the same form (of course with the ballistic fluctuation propagator) such that the $1/(i\omega)$ term \eqref{eq:divergent1} vanishes identically. This fact is intimately related with the gauge invariance (see Ref. \cite{Anderson2016,Boyack} for detailed discussion). Indeed the expression \eqref{eq:divergent1} can be written as 
the second derivative of the contribution to the thermodynamic potential from superconducting fluctuations with respect to a constant vector potential. Since the thermodynamic potential is independent of the constant vector potential in virtue of the gauge invariance, the expression \eqref{eq:divergent1} should be zero. We note that the result for quantum correction to the ac conductivity due to interaction in the Cooper channel reported in Ref. \cite{PV} diverges as $1/(i\omega)$ in the limit $\omega \to 0$. 

Gathering together the contributions \eqref{app:MT:an:t}, \eqref{app:MT:reg:t:final}, \eqref{app:DOS:t:final}, and \eqref{app:AL:t0} (disregarding the terms which sum up to zero as discussed above), we find the following final form of the correction to the ac conductivity due to the interaction in the Cooper channel:
\begin{equation}
\delta g^{CC}(\omega) = \delta g^{\rm MT, an}(\omega) + \delta g^{\rm DOS}(\omega) + \delta g^{\rm AL}(\omega) + \delta g^{\rm sc}(\omega) .
\label{eq:CC:f}
\end{equation}
Here we introduce $\delta g^{\rm sc}(\omega) =\delta g^{\rm sc, 1}(\omega)+\delta g^{\rm sc, 2}(\omega)+\delta g^{\rm sc, 3}(\omega)$ that can be rewritten as the following lengthy expression:
\begin{gather}
\delta {g}^{\rm sc}(\omega)
=\frac{D}{4\pi d T \omega } \int_{q,\Omega} 
\partial_{q_\mu} \Biggl\{  q_\mu \mathcal{L}_q^R(\Omega) \Biggl [ 4 \mathcal{B}_\Omega \Phi_{-2\omega}(\Omega)
+ \mathcal{B}_\Omega
\Bigl[ \psi^\prime(\mathcal{X}_{q,\Omega})-\psi^\prime(\mathcal{X}_{q,\Omega+2\omega})
\Bigr]
+  \Bigl[ \mathcal{B}_\Omega-\mathcal{B}_{\Omega-\omega}\Bigr ]
\notag\\
\times \Bigl[ \psi^\prime(\mathcal{X}_{q,\Omega})
-\psi^\prime(\mathcal{X}_{q,2\omega-\Omega})
\Bigr] \Biggr ] \Biggr\}
+ \frac{D^2}{8d (\pi T)^2 \omega } \int_{q,\Omega} q^2 
\mathcal{L}_q^R(\Omega) \mathcal{B}_\Omega \Biggl [
3 \psi^{\prime\prime}(\mathcal{X}_{q,\Omega})
+ \psi^{\prime\prime}(\mathcal{X}_{q,\Omega+2\omega})
\notag \\
+2 \left (\frac{4\pi T}{\omega}\right)^2 \Bigl [ \psi(\mathcal{X}_{q,2\omega+\Omega}) - \psi(\mathcal{X}_{q,\Omega+\omega})
- \psi(\mathcal{X}_{q,\Omega}) + \psi(\mathcal{X}_{q,\Omega-\omega})
\Bigl ]
\Biggl ]
+\frac{D^2}{8d (\pi T)^2 \omega } \int_{q,\Omega} q^2 
\mathcal{L}_q^R(\Omega) 
\notag \\
\times
\Bigl[ \mathcal{B}_{\Omega-\omega}-\mathcal{B}_{\Omega}\Bigr ]
\Biggl [ \psi^{\prime\prime}(\mathcal{X}_{q,2\omega-\Omega})-\psi^{\prime\prime}(\mathcal{X}_{q,\Omega})-\frac{8\pi T}{i\omega} \Bigl [ \Phi_\omega(\Omega)-\Phi_\omega(2\omega-\Omega)\Bigr]
\Biggr ]
\notag\\
- \frac{D^2}{8 d(\pi T)^2\omega} \int_{q,\Omega} q^2 
\mathcal{L}_q^R(\Omega) \mathcal{B}_\Omega
\Biggl \{\mathcal{L}_q^R(\Omega) \psi^{\prime}(\mathcal{X}_{q,\Omega}) 
\Bigl [ 4 \Phi_{-2\omega}(\Omega) 
- \psi^{\prime}(\mathcal{X}_{q,\Omega+2\omega})- 
7 \psi^{\prime}(\mathcal{X}_{q,\Omega}) \Bigr ]
\notag \\
+2 \mathcal{L}_q^R(\Omega-\omega) \bigl [ \Delta^{RRR}_q(\Omega-\omega,\Omega,\omega)\bigr ]^2 \Biggr \}
\notag \\
- \frac{D^2}{8 d(\pi T)^2\omega} \int_{q,\Omega} q^2 
\mathcal{L}_q^R(\Omega) \Bigl [\mathcal{B}_\Omega- 
\mathcal{B}_{\Omega-\omega} \Bigr ]
\Biggl \{
\mathcal{L}_q^R(\Omega) 
\psi^{\prime}(\mathcal{X}_{q,\Omega})
\Bigl [\psi^{\prime}(\mathcal{X}_{q,\Omega})
- \psi^{\prime}(\mathcal{X}_{q,2\omega-\Omega})
 \Bigr ]
 \notag \\
 - \mathcal{L}_q^R(\Omega-\omega)\Bigl [
 \bigl ( \Delta^{RRR}_q(\Omega-\omega,\Omega,\omega)\bigr )^2
 +i \Delta^{RRR}_q(\Omega-\omega,\Omega,\omega) \im \Delta^{RRR}_q(\Omega-\omega,\Omega,\omega)\Bigr ]
 \notag \\
 - \mathcal{L}_q^A(\Omega-\omega)\Bigl [
 \Delta^{RRR}_q(\Omega-\omega,\Omega,\omega) \re \Delta^{RRR}_q(\Omega-\omega,\Omega,\omega) -\bigl ( \Delta^{ARR}_q(\Omega-\omega,\Omega,\omega)\bigr )^2\Bigr ]
\Biggr \} .
\label{eq:sigma:sc:f}
\end{gather}
We note that the first term in the right hand side of Eq. \eqref{eq:sigma:sc:f} is the full derivative with respect to momentum and, thus, as discussed above should vanish being supplemented by the corresponding contribution from the ballistic scales. Therefore, we shall disregard the corresponding term below.

\subsection{Corrections to the conductivity in the dc limit due to superconducting fluctuations}

Although corrections to the static conductivity due to superconducting fluctuations were discussed many times in literature, it is instructive to check that our result \eqref{eq:CC:f} for an arbitrary frequency correctly reproduces the well-known corrections in the static limit. In particular, the static anomalous Maki--Thompson correction becomes
\begin{equation}
\delta g^{\rm MT, an}(\omega=0) =
 4 \int_{q,\Omega} \mathcal{C}_q^R(0) \partial_\Omega \mathcal{B}_{\Omega} \im \mathcal{L}_q^A(\Omega) \im \psi(\mathcal{X}_{q,\Omega}) .
\label{app:MT:an:final:dc}
\end{equation}
The DOS correction in the dc limit, $\omega\to 0$, acquires the following form 
\begin{align}
\delta g^{\rm DOS}(\omega=0) = -\frac{D}{8\pi^2 T^2} \im \int_{q,\Omega} \mathcal{B}_\Omega
\mathcal{L}^R_q(\Omega) \psi^{\prime\prime}\left (\mathcal{X}_{q,\Omega}\right ) - \frac{D}{2\pi T} \int_{q,\Omega} \partial_\Omega \mathcal{B}_\Omega \im\mathcal{L}^R_q(\Omega) \im\psi^{\prime}(\mathcal{X}_{q,\Omega}) .
\label{eq_Sigma_DOS_dc}
\end{align}
At $\omega\to 0$ the Aslamazov--Larkin correction can be written as follows
\begin{align}
\delta{g}^{\rm AL}(\omega=0) =  - \frac{D^2}{d\pi^2 T^2} \int_{q,\Omega} q ^2  \partial_\Omega \mathcal{B}_{\Omega}
\im \mathcal{L}_q^R(\Omega)\im \Bigl[\mathcal{L}_q^R(\Omega) \psi^{\prime}(\mathcal{X}_{q,\Omega})\Bigr ]  \re \psi^{\prime}(\mathcal{X}_{q,\Omega})
 .
\label{sigma:AL:dc}
\end{align}
Finally, the contribution $\delta g^{\rm sc}$ in the dc limit becomes
\begin{align}
\delta g^{\rm sc}(\omega=0) & = - \frac{D^2}{2d(2\pi T)^3} \im \int_{q,\Omega} q^2
\mathcal{B}_\Omega 
\mathcal{L}^{R2}_q(\Omega)\psi^{\prime}\left (\mathcal{X}_{q,\Omega}\right ) \psi^{\prime\prime}\left (\mathcal{X}_{q,\Omega}\right ) 
\notag \\
& - \frac{D^2}{d(2\pi T)^2} \int_{q,\Omega} q^2  \partial_\Omega \mathcal{B}_\Omega  \im \Bigl[ \mathcal{L}^{R2}_q(\Omega)\psi^{\prime}\left (\mathcal{X}_{q,\Omega}\right )\Bigr ] 
\im\psi^{\prime}\left (\mathcal{X}_{q,\Omega}\right ) .
\label{eq:sigma:sc:dc}
\end{align}
We note that Eqs. \eqref{app:MT:an:final:dc}, \eqref{eq_Sigma_DOS_dc}, \eqref{sigma:AL:dc}, and \eqref{eq:sigma:sc:dc} coincide with the zero magnetic field limit of corresponding fluctuation corrections found in Ref. \cite{KSF2012} and with the fluctuation corrections in the diffusive regime computed in Ref. \cite{StepanovSkvortsov}.

\section{Corrections to the ac conductivity due to superconducting fluctuations}
\label{s4}

Now we discuss the dependence of corrections to the ac conductivity due to superconducting fluctuations. It is convenient to introduce the following dimensionless variables, $\epsilon = \ln T/T_c$ and $\alpha = \omega/(4\pi T)$. 

\subsection{Anomalous Maki--Thompson contribution}

We start from the anomalous Maki--Thompson correction, Eq. \eqref{app:MT:an:t}. We note that the integral over momentum in Eq. \eqref{app:MT:an:t} diverges in the infra-red. Therefore, we need to introduce a finite dephasing rate $1/\tau_\phi$ which cuts off the pole in the cooperon propagator. In what follows we shall use dimensionless variable $\gamma = 1/(4\pi T\tau_\phi)$.

The asymptotic behavior of $\delta g^{\rm MT, an}(\omega)$  at large frequencies,
$\alpha\gg 1$, and for an arbitrary distance from superconducting transition temperatue, $\epsilon$, is given as follows (see \ref{App:MTan})
\begin{gather}
\delta g^{\rm MT, an}(\omega) = \frac{\pi^2-8\ln 2}{4\pi} \frac{1}{\epsilon+\ln \alpha} 
.
\label{eq:g:Mtan:large:f}
\end{gather}
The anomalous Maki--Thompson correction vanishes in the limit of large frequencies, $\omega \gg T$. Away from the superconducting transition, $\epsilon \gg 1$, and for small frequencies,
$\alpha\ll 1$, the anomalous Maki--Thompson contribution becomes 
\begin{gather}
\delta g^{\rm MT, an}(\omega) = \left (\frac{\pi}{6\epsilon^2}-  \frac{2\pi i  \alpha}{3 \epsilon}
 \right ) \ln \frac{1}{\gamma-i\alpha} . 
\label{eq:MTan:regII}
\end{gather}
We note that the first term in the r.h.s. of Eq. \eqref{eq:MTan:regII} dominates over the second one at $\alpha \ll 1/\epsilon$. At small frequencies, $\alpha\ll 1$, and in the vicinity of the superconducting transition, $\epsilon \ll 1$, the anomalous Maki-Thompson correction reads
\begin{gather}
\delta g^{\rm MT, an}(\omega) = \frac{1}{2\pi} \frac{1}{\bar\epsilon-\gamma+i\alpha}\ln 
\frac{\bar\epsilon}{\gamma-i\alpha} ,
\label{eq:MTan:regI}
\end{gather}
where $\bar\epsilon=2\epsilon/\pi^2 \equiv 1/(4\pi T\tau_{GL})$. We note that  we omit subleading terms proportional to $\ln \epsilon$ in Eq. \eqref{eq:MTan:regI} (see Refs. \cite{PV,StepanovSkvortsov} for details).

\begin{figure}[t]
\centerline{\includegraphics[width=0.5\textwidth]{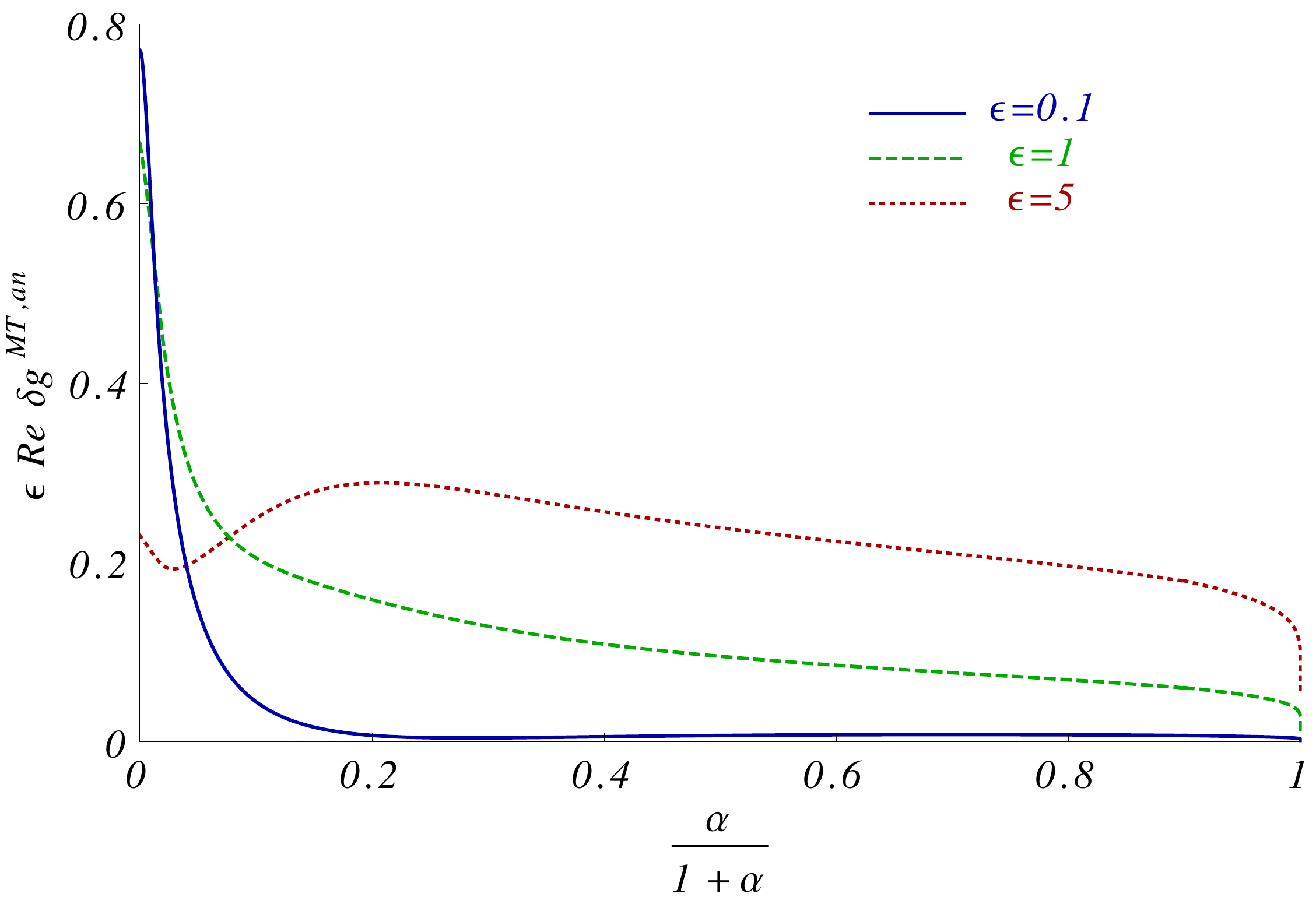}\quad \includegraphics[width=0.51\textwidth]{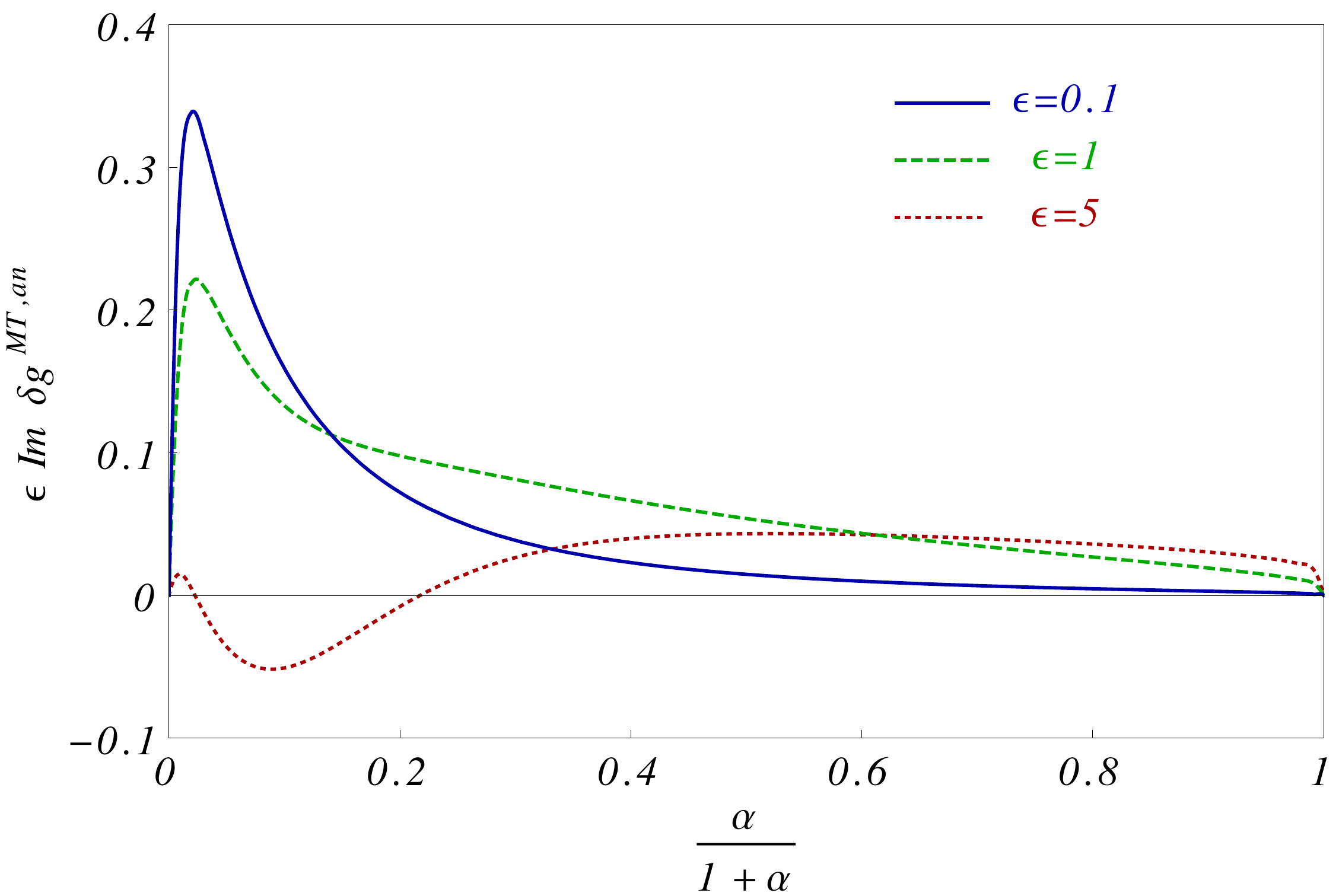}}
\caption{The dependence of the real (left panel) and imaginary (right panel) parts of the anomalous Maki-Thompson correction on the frequency at different temperatures. The ratio of the dephasing rate to the temperature is fixed to the value $\gamma=0.01$.}
\label{Figure1}
\end{figure}

The overall behavior of $\delta g^{\rm MT, an}(\omega)$ as a function of the dimensionless frequency $\alpha$ at different values of $\epsilon$ is shown in Fig. \ref{Figure1}.  The real part of 
$\delta g^{\rm MT, an}(\omega)$ has non-monotonous behavior for temperatures close to $T_c$, i.e. for $\epsilon \ll 1 $ (see the left panel in Fig. \ref{Figure1}). For temperatures away from $T_c$, i.e. for $\epsilon \gg 1$, $\re \delta g^{\rm MT, an}(\omega)$ is also non-monotonous function of $\omega$. In the case $T\gg T_c$, provided $1/\tau_\phi\ll T/\ln (T/T_c)$, the real part of  $\delta g^{\rm MT, an}(\omega)$ has the minimum at $\omega \sim T/\ln (T/T_c)$. 
The dependence of the imaginary part of $\delta g^{\rm MT, an}(\omega)$ 
on the dimensionless frequency $\alpha$ at different values of $\epsilon$ is figured in the right panel of Fig. \ref{Figure1}. Exactly at zero frequency the imaginary part vanishes, $\im \delta g^{\rm MT, an}(\omega=0)=0$. The imaginary part of $\delta g^{\rm MT, an}(\omega)$ demonstrates non-monotonous behavior with $\omega$.
At ultra small frequencies, $\omega\ll 1/\tau_\phi$, the imaginary part of $\delta g^{\rm MT, an}(\omega)$ increases linearly with $\omega$. For $T\gg T_c$, $\im \delta g^{\rm MT, an}(\omega)$ has the maximum at the frequency of the order of $\sqrt{(T/\tau_\phi)/\ln (T/T_c)}$. For temperatures near $T_c$, i.e. for $\epsilon\ll 1$, the imaginary part of $\delta g^{\rm MT, an}(\omega)$ has the maximum at $\omega \sim 1/\sqrt{\tau_\phi\tau_{GL}}$.

\subsection{DOS correction}

Next we turn our attention to the DOS correction to the conductivity. We note that the integrals over momentum and frequency in Eq. \eqref{eq_Sigma_DOS}
diverge at the ultraviolet. Therefore, we shall introduce a cut-off corresponding to the inverse elastic mean free time, $1/\tau$. Then, we can single out the part of $\delta g^{\rm DOS}(\omega)$  that depends on the cut-off, 
\begin{equation}
\delta g^{\rm DOS}(\omega) =-\frac{1}{\pi} \ln \ln \frac{1}{4\pi T_c\tau} +  \delta g^{\rm DOS}_{f}(\omega) ,
\label{eq:DOS:split}
\end{equation}
such that $\delta g^{\rm DOS}_{f}(\omega)$ is finite in the ultraviolet.
We mention that the first term in the right hand side of Eq. \eqref{eq:DOS:split} corresponds to the one loop DOS correction in the renormalization group equations for the conductivity \cite{Fin1984Z}.

\begin{figure}[t]
\centerline{\includegraphics[width=0.5\textwidth]{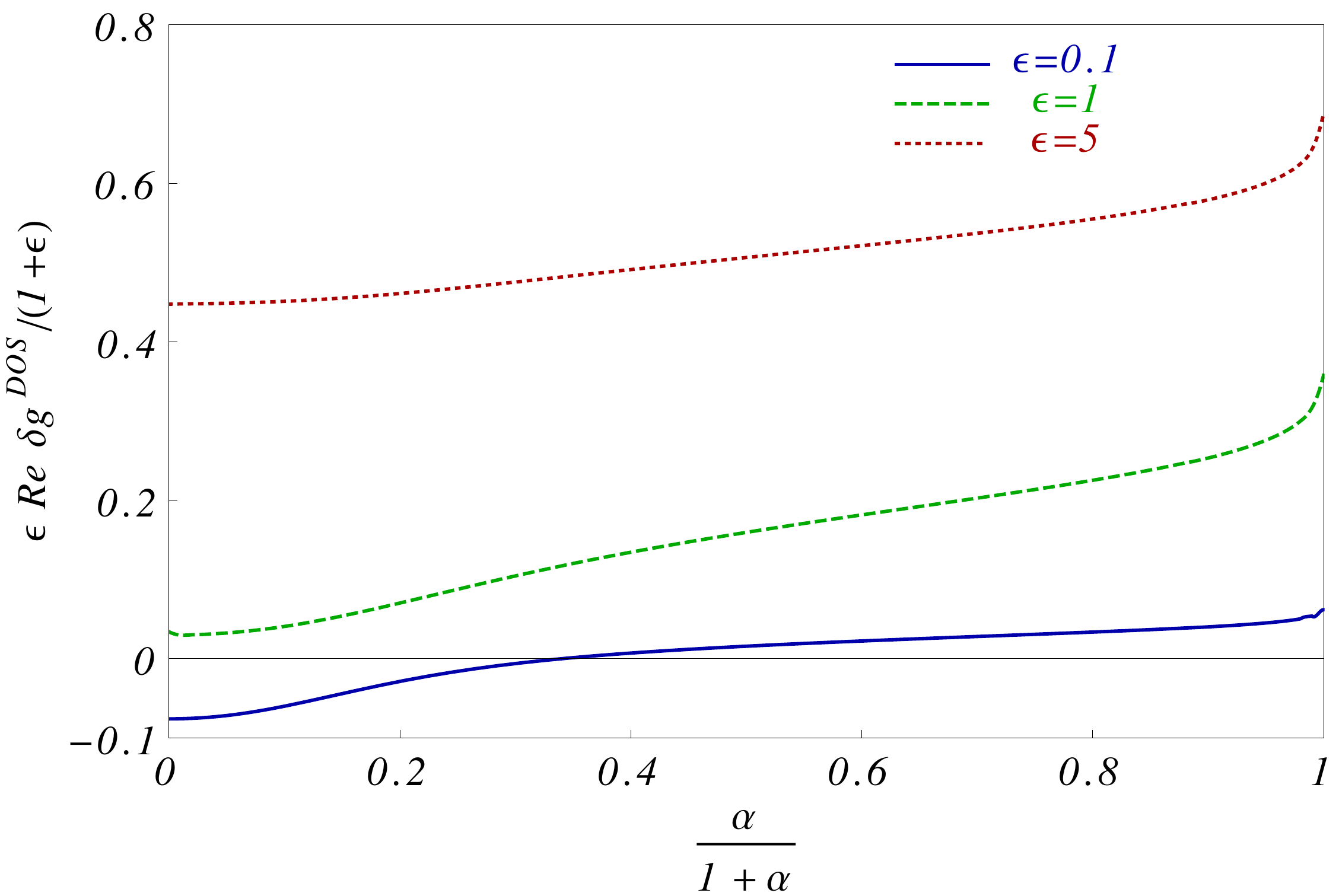}\quad \includegraphics[width=0.51\textwidth]{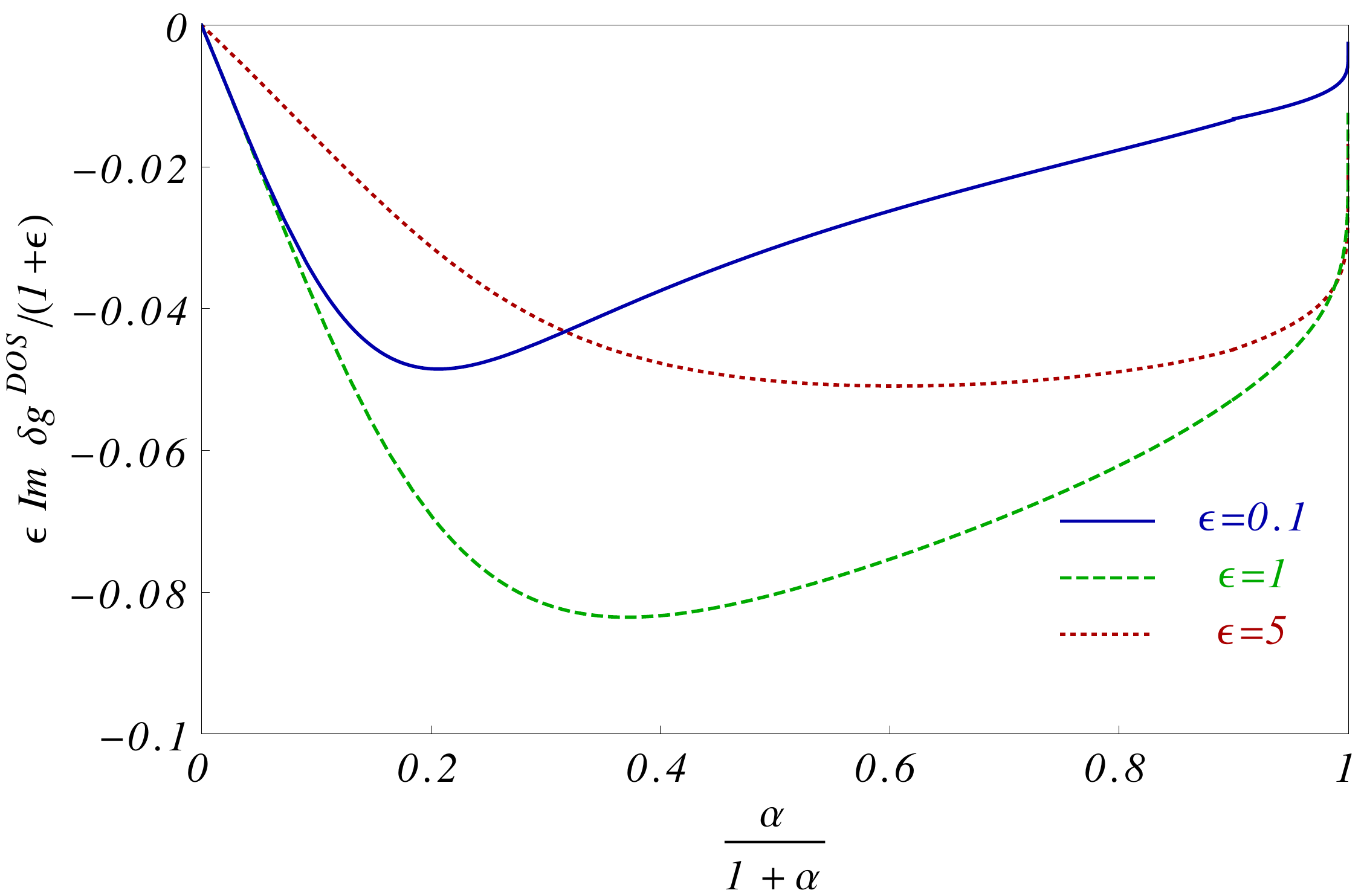}}
\caption{The dependence of the real (left panel) and imaginary (right panel) parts of the DOS correction on the frequency at different temperatures.}
\label{Figure2}
\end{figure}

At large frequencies, $\alpha\gg 1$, and for an arbitrary magnitude of $\epsilon$ the asymptotic behavior of $\delta g^{\rm DOS}_f(\omega)$ is given as  (see \ref{App_DOS})
\begin{gather}
\delta g^{\rm DOS}_f(\omega)  = \frac{1}{\pi} \ln \Bigl(\epsilon+\ln \alpha \Bigr )-\frac{i}{2} \frac{1}{\epsilon+\ln \alpha} .
\label{eq:DOS:large:f}
\end{gather}
In the case of high temperatures, $\epsilon\gg 1$, but small frequencies, $\alpha\ll 1$, the DOS correction becomes
\begin{gather}
\delta g^{\rm DOS}_f(\omega)  = \frac{1}{\pi} \ln \epsilon +\frac{\ln 2}{\pi \epsilon} - \frac{2 \pi i \alpha}{3\epsilon}.
\label{eq:DOS:large:regII}
\end{gather}
Near the superconducting transition, $\epsilon\ll 1$, and at small frequencies, $\alpha\ll 1$, one can derive the following expression
\begin{gather}
\delta g^{\rm DOS}_f(\omega)  = -\left (\frac{14 \zeta(3)}{\pi^3}+ i\pi\alpha \right )\ln \frac{1}{\epsilon} .
\label{eq:DOS:large:regIII}
\end{gather}

The overall dependence of the real and imaginary parts of $\delta g^{\rm DOS}_f(\omega)$ on frequency is shown in Fig. \ref{Figure2}. The real part of $\delta g^{\rm DOS}_f(\omega)$ grows monotonically with increase of the frequency. The imaginary part has the minimum.

\subsection{Aslamazov--Larkin contribution}

Next we consider the Aslamazov--Larkin contribution to the conductivity. We note that this correction is finite both in the infrared and the ultraviolet. We start from the case of  large frequencies, $\alpha\gg 1$, and arbitrary temperature above $T_c$. Then we find  (see \ref{App:AL:Sec})
\begin{gather}
\delta g^{\rm AL}(\omega)
= \frac{c^{\rm AL}_3}{(\epsilon+\ln\alpha)^3}, 
\label{eq:AL:large:f}
\end{gather}
where numerical constant $c^{\rm AL}_3 \approx 0.17 - 0.89 i$. In the case of small frequencies, $\alpha\ll 1$, and temperatures away from the superconducting transition, $\epsilon\gg 1$,
we obtain
\begin{gather}
\delta g^{\rm AL}(\omega) 
= \frac{c^{\rm AL}_4- i c^{\rm AL}_5 \alpha}{\epsilon^3} ,
\label{eq:AL:large:regII}
\end{gather}
where magnitudes of the numerical constants are $c^{\rm AL}_4\approx 1.44$ and $c^{\rm AL}_5\approx 9.23$. For temperatures close to superconducting transitions, $\epsilon\ll 1$, and for small frequencies, $\alpha\ll 1$, the Aslamazov--Larkin contribution becomes
\begin{gather}
\delta g^{\rm AL}(\omega) 
= \frac{\pi}{8 \epsilon} W_1\left (\frac{\pi^2\alpha}{2\epsilon}\right ) 
- \frac{i \pi^3\alpha}{32\epsilon^2}  W_2\left (\frac{\pi^2\alpha}{2\epsilon}\right ) .
\label{eq:AL:fl}
\end{gather}
Here the functions $W_{1,2}(z)$ are defined as follows 
\begin{gather}
W_1(z) = \frac{4}{z}\Bigl [\arctan(z/2)-\frac{1}{z} \ln(1+z^2/4)\Bigr ],\notag \\
W_2(z) = \frac{8}{z^3} \Bigl [\arctan(z)-2 \arctan(z/2)+ z\arctan\frac{3z^2}{8+5z^2}\Bigr ] .
\label{eq:W12}
\end{gather}
We note that the part of Eq. \eqref{eq:AL:fl} proportional to the function $W_1$ coincides with the result derived in Ref. \cite{Schmidt} and with the contribution $\re \sigma^{(1,1)}_{AL}$ of Ref. \cite{PV}. We note that there is also subleading term proportional to $\ln \epsilon$ (see Eq. (41) of Ref. \cite{PV}).

\begin{figure}[t]
\centerline{\includegraphics[width=0.5\textwidth]{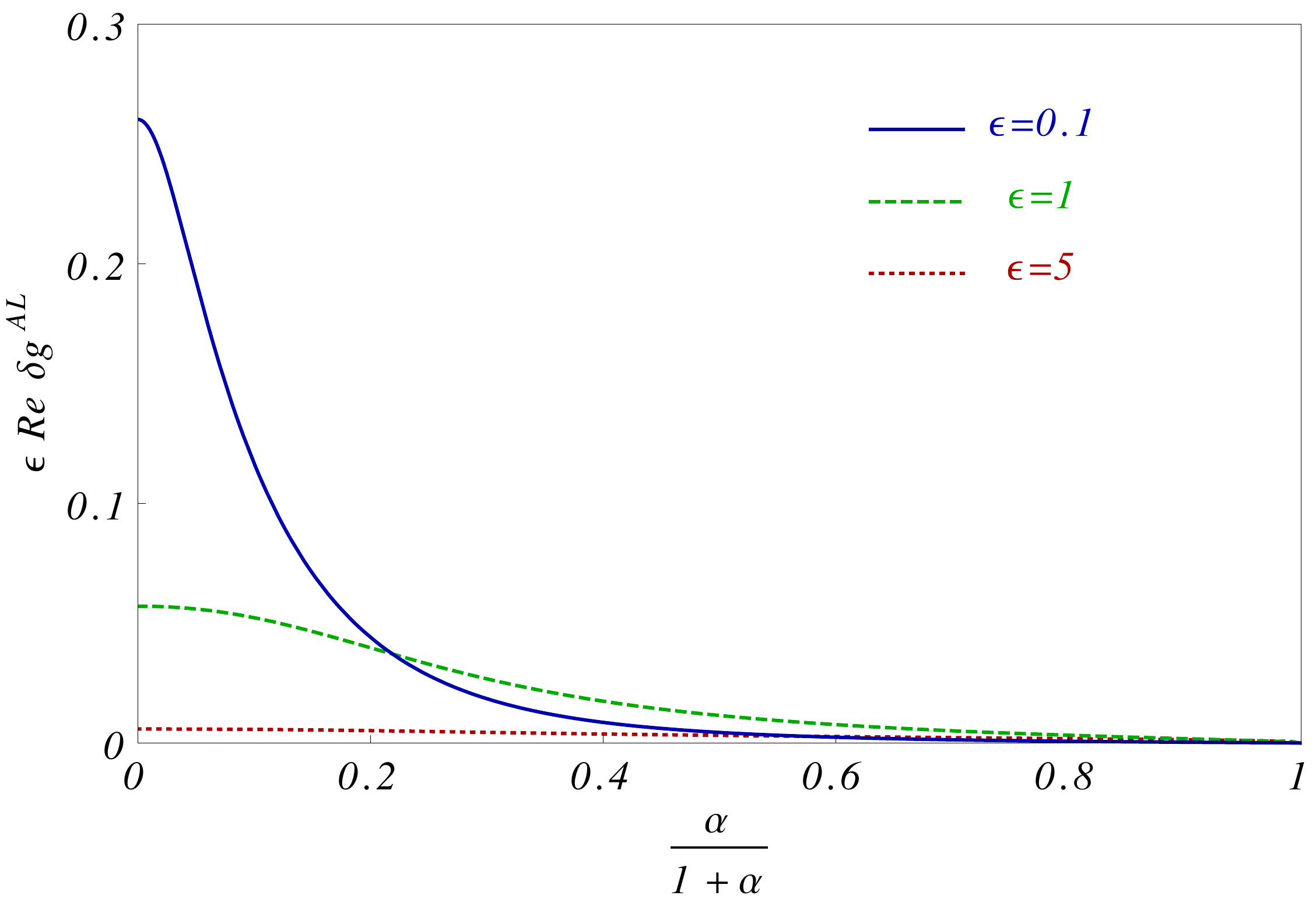}\quad \includegraphics[width=0.51\textwidth]{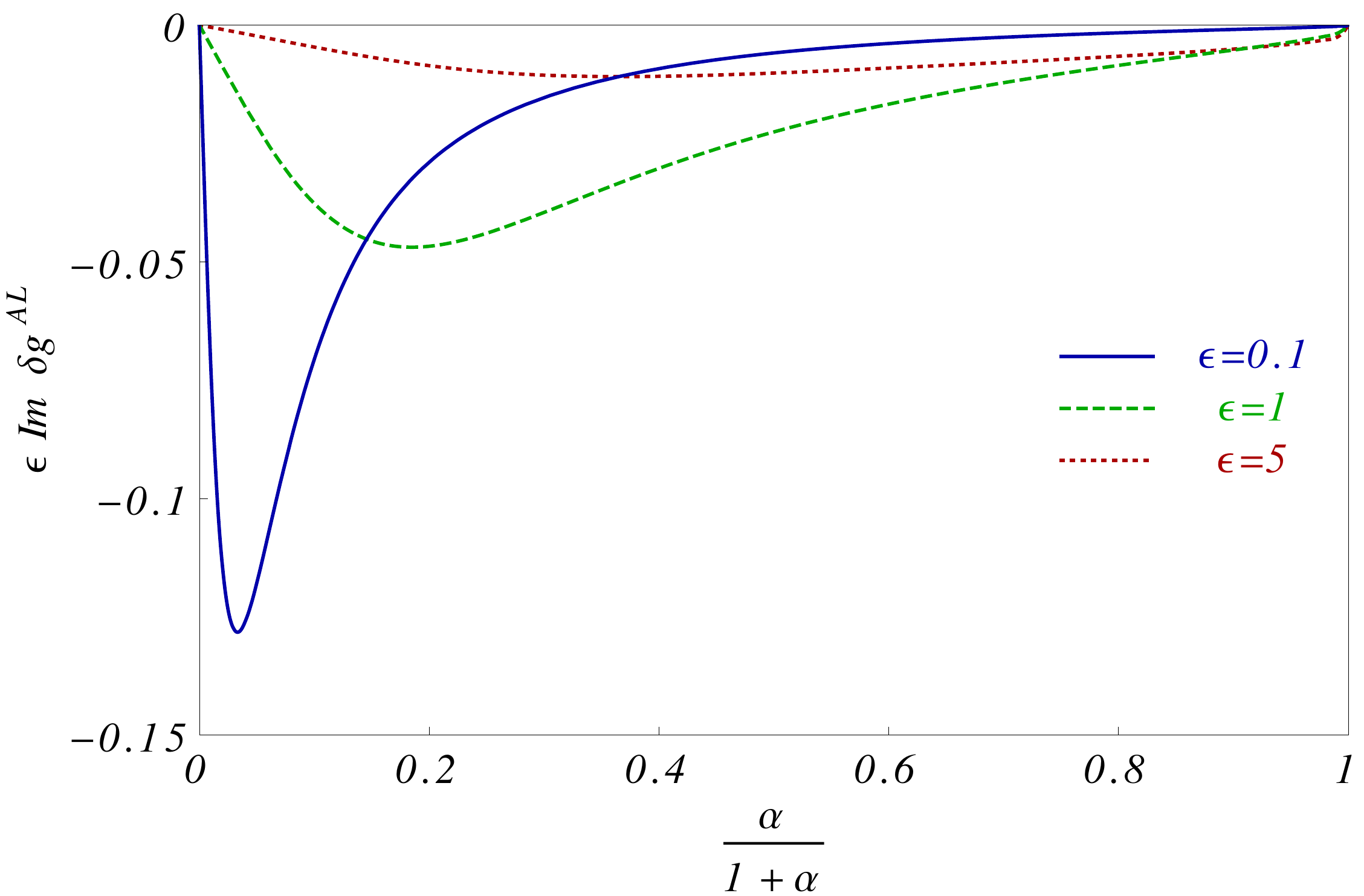}}
\caption{The dependence of the real (left panel) and imaginary (right panel) parts of the Aslamazov-Larkin correction on the frequency at different temperatures.}
\label{Figure3}
\end{figure}

The overall dependence of the real and imaginary parts of $\delta g^{\rm AL}(\omega)$ on frequency is shown in Fig. \ref{Figure3}. The real part of $\delta g^{\rm AL}(\omega)$ decreases monotonously with increase of  $\alpha$. The imaginary part of $\delta g^{\rm AL}(\omega)$ has the minimum at some frequency for all temperatures above the superconducting transition. For $T$ close to $T_c$ the maximum is at $\alpha \sim \epsilon$. 

\subsection{The correction $\delta g^{\rm sc}(\omega)$}

Finally, we turn our attention to the contribution  $\delta g^{\rm sc}(\omega)$, cf. Eq. \eqref{eq:sigma:sc:f}. Similar to the Aslamasov--Larkin contribution, the correction $\delta g^{\rm sc}(\omega)$ has divergencies neither in the infrared nor in the ultraviolet. At first, we consider the case of  large frequencies, $\alpha\gg 1$, and arbitrary temperatures above the superconducting transition. Then we find  (see \ref{App:gsc:Sec})
\begin{gather}
\delta g^{\rm sc}(\omega) = \frac{2}{3\pi} \frac{1+3\ln 2}{\epsilon+\ln\alpha} .
\label{eq:as:sc:1}
\end{gather}
In the case of small frequencies, $\alpha\ll 1$, but high temperatures, $\epsilon\gg 1$, we obtain
\begin{gather}
\delta g^{\rm sc}(\omega) = \frac{1}{2\pi \epsilon} \left (1-\frac{14\pi^2 i \alpha}{3}\right ) .
\label{eq:as:sc:2}
\end{gather}
In the vicinity of the superconducting transition, $\epsilon\ll 1$, and small frequencies, $\alpha\ll 1$, we find 
\begin{gather}
\delta g^{\rm sc}(\omega) = \frac{i \pi^3 \alpha}{24\epsilon^2} W_3\left (\frac{\pi^2\alpha}{2\epsilon}\right )
- \frac{28 \zeta(3)}{\pi^3} \ln \epsilon  ,
\label{eq:as:sc:3}
\end{gather}
where
\begin{equation}
W_3(z) = \frac{3}{z^2} \left [ 2 \frac{\arctan z}{z} + \ln(1+z^2)-2\right ] .
\label{eq:W3:def}
\end{equation}

\begin{figure}[t]
\centerline{\includegraphics[width=0.5\textwidth]{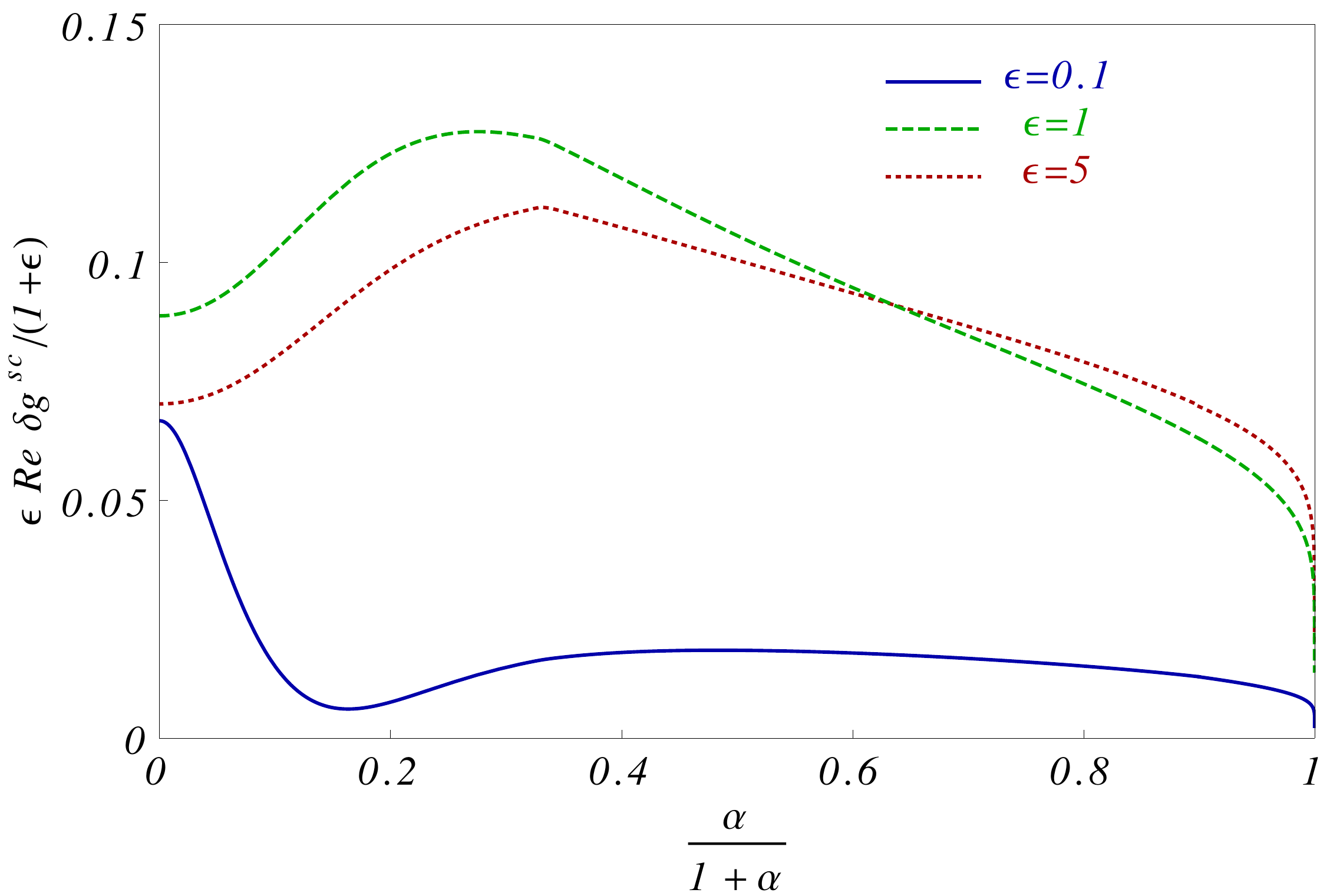}\quad \includegraphics[width=0.51\textwidth]{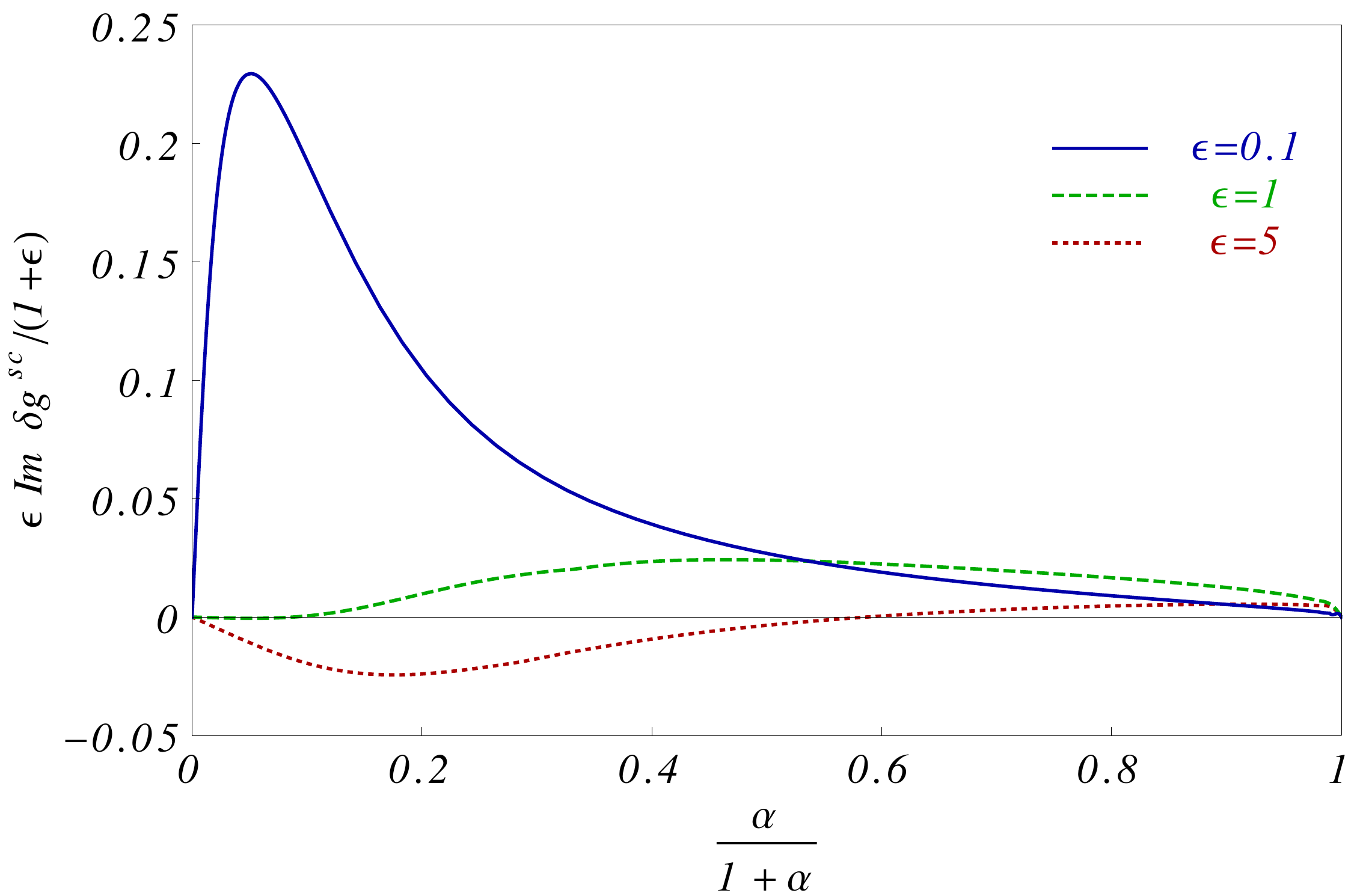}}
\caption{The dependence of the real (left panel) and imaginary (right panel) parts of  $\delta g^{\rm sc}(\omega) $ on the frequency at different temperatures.}
\label{Figure4}
\end{figure}

The overall dependence of the real and imaginary parts of $\delta g^{\rm sc}(\omega)$ on frequency is shown in Fig. \ref{Figure4}. Both the real and imaginary parts of $\delta g^{\rm sc}(\omega)$ 
have non-monotonous behavior. For temperatures away from the superconducting transition, $\epsilon\gg 1$, $\re \delta g^{\rm sc}(\omega)$ is positive and has the maximum at some frequency $\alpha \sim 1$. The imaginary part of $\delta g^{\rm sc}(\omega)$ has the minimum at some frequency $\alpha$ of the order of unity.
Near the superconducting transition, $\epsilon\ll 1$, 
the real (imaginary) part of $\delta g^{\rm sc}(\omega)$ is positive and has the minimum (maximum) at $\alpha\sim \epsilon$. 

\subsection{The asymptotic expressions for $\delta g^{CC}(\omega)$}

Now we are ready to present the asymptotic expressions for the correction to the ac conductivity due to superconducting fluctuations, i.e. due to interaction in the Cooper channel. It is convenient to single out the term which depends on the ultraviolet cutoff $1/\tau$,
\begin{gather}
\delta g^{CC}(\omega)  = -\frac{1}{\pi} \ln \ln[1/(4\pi T_c\tau)]+ \delta g^{CC}_f(\omega) .
\end{gather}
The contribution $\delta g^{CC}_f(\omega)$ is finite in the ultraviolet. For large frequencies  in comparison with the temperature, $\omega \gg T$, we find from Eqs. \eqref{eq:g:Mtan:large:f}, \eqref{eq:DOS:large:f},  \eqref{eq:AL:large:f}, and \eqref{eq:as:sc:1}, 
\begin{gather}
\delta g^{CC}_f(\omega)  = \frac{1}{\pi} \ln \ln [\omega/(4\pi T_c)] + \frac{3\pi^2+8- 6 \pi i}{12 \pi \ln [\omega/(4\pi T_c)]}  .
\end{gather}
As expected the real and imaginary parts of the conductivity correction is dominated by the DOS contribution, Eq. \eqref{eq:DOS:large:f}. At small frequencies, $\omega \ll T$, but for temperatures away from the superconducting transition, $T\gg T_c$, using Eqs. \eqref{eq:MTan:regII}, \eqref{eq:DOS:large:regII}, \eqref{eq:AL:large:regII}, and 
\eqref{eq:as:sc:2}, we obtain
\begin{gather}
\delta g^{CC}_f(\omega)  =\frac{1}{\pi} \ln \ln (T/T_c) + \frac{2\ln 2+1}{2\pi \ln (T/T_c)} 
+ \frac{1}{6} \left (\frac{i\omega}{T}-\frac{\pi}{\ln(T/T_c)}\right ) \frac{\ln[(\tau_\phi^{-1}-i\omega)/(4\pi T)]}{\ln(T/T_c)} .
\end{gather}
The real part of the conductivity correction is dominated by the DOS contribution as in the static case. The imaginary part of the conductivity correction is dominated by the anomalous Maki--Thompson term. In the region close to the superconducting transition, $T-T_c\ll T_c$, and for small frequencies, $\omega \ll T_c$, with the help of Eqs. \eqref{eq:MTan:regI}, \eqref{eq:DOS:large:regIII}, \eqref{eq:AL:fl}, and \eqref{eq:as:sc:3}, we find
\begin{gather}
\delta g^{CC}_f(\omega) = - \frac{2 T\tau_{GL} }{1-\tau_{GL}/\tau_\phi+i\omega\tau_{GL}}\ln 
\Bigl[ \tau_{GL}/\tau_\phi-i\omega\tau_{GL}\Bigr ]
+ T\tau_{GL} \Bigl [W_1(\omega  \tau_{GL}) - \frac{i\omega  \tau_{GL}}{2}W_2(\omega  \tau_{GL})\notag \\
+\frac{i\omega  \tau_{GL}}{3}W_3(\omega  \tau_{GL})\Bigr ] .
\end{gather}

\begin{figure}[t]
\centerline{\includegraphics[width=0.5\textwidth]{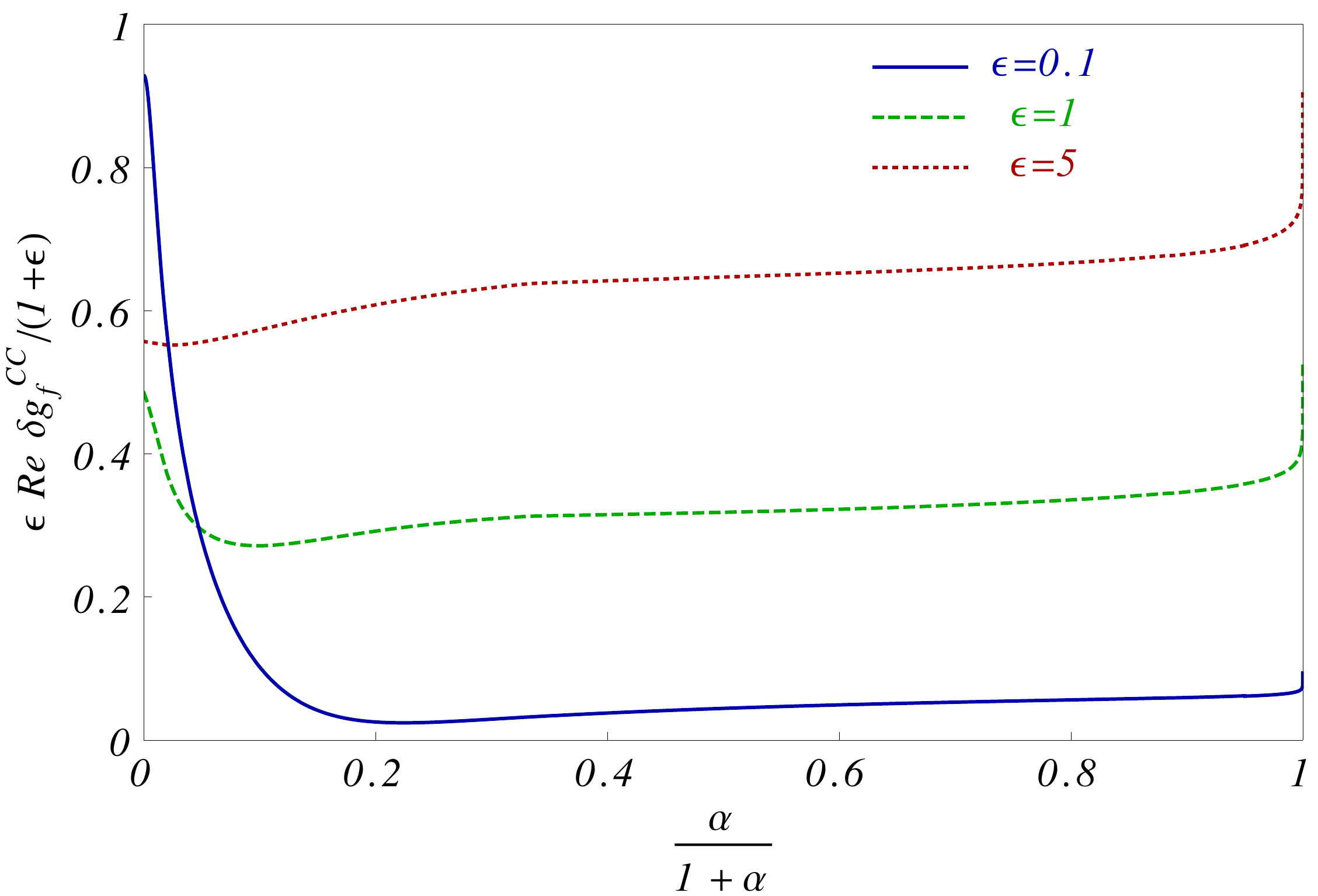}\quad \includegraphics[width=0.5\textwidth]{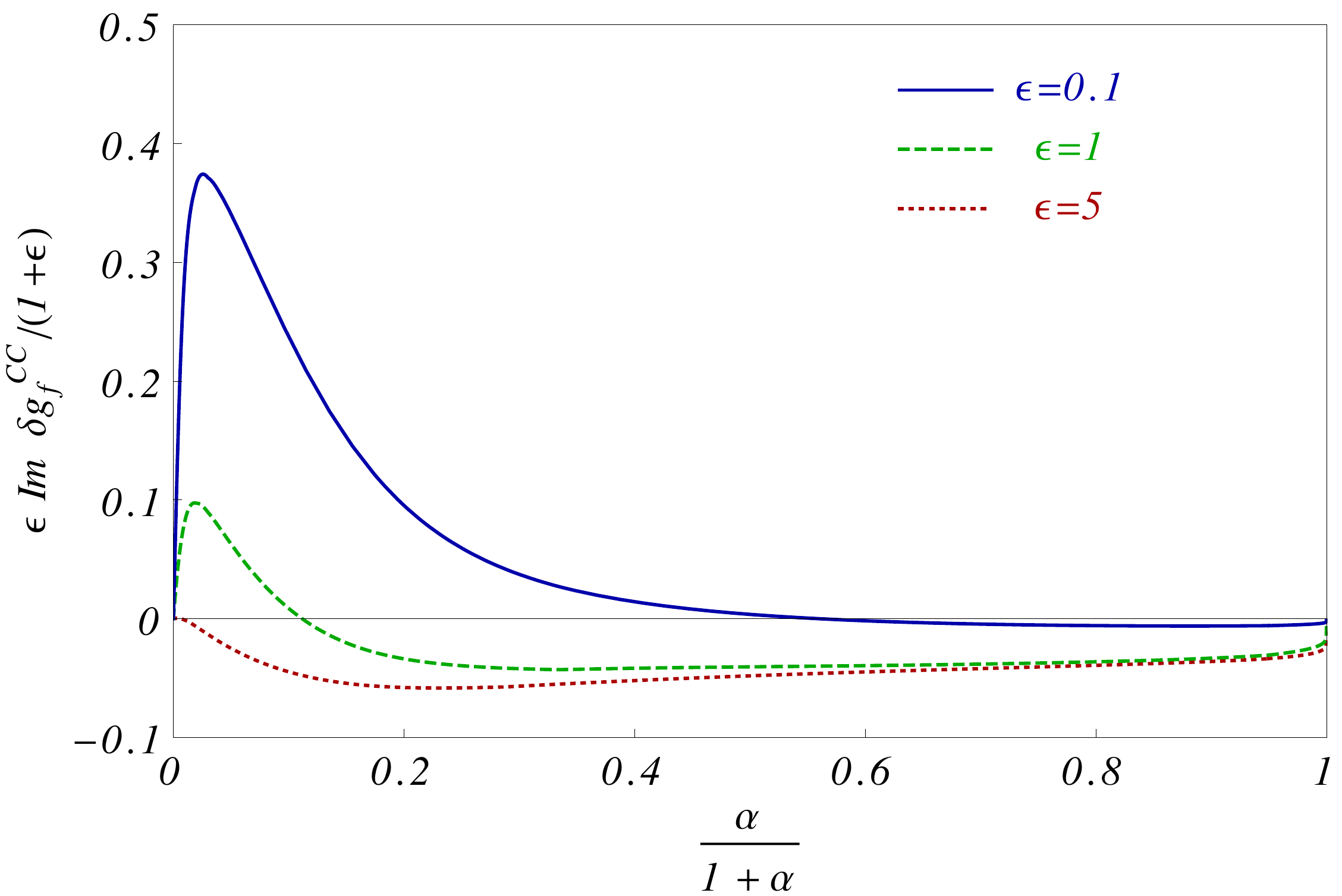}}
\caption{The dependence of the real (left panel) and imaginary (right panel) parts of  $\delta g^{CC}_f(\omega)$ on the frequency at different temperatures. The ratio of the dephasing rate to the temperature is fixed to the value $\gamma=0.01$.}
\label{Figure5}
\end{figure}

The dependence of real and imaginary parts of $\delta g^{CC}_f(\omega)$ on frequency for different temperatures is shown in Fig. \ref{Figure5}. For all temperatures above  $T_c$ the real (imaginary) part of $\delta g^{CC}_f(\omega)$ has the minimum (maximum). At temperatures $T\gg T_c$, the minimum of $\re\delta g^{CC}_f(\omega)$ 
occurs at frequency $\omega \sim 1/\ln (T/T_c)$ whereas the maximum of $\im\delta g^{CC}_f(\omega)$ is at $\omega \sim \sqrt{T/[\tau_\phi \ln(T/T_c)]}$. In the vicinity of the superconducting transition, $T-T_c\ll T_c$, the  real part of $\delta g^{CC}_f(\omega)$ has a shallow minimum at frequency of the order of $T_c$. The maximum of the imaginary part of $\delta g^{CC}_f(\omega)$ is at frequency $\omega \sim 1/\sqrt{\tau_\phi\tau_{GL}}$. The frequency dependence of the real and imaginary part of $\delta g^{CC}_f(\omega)$ shown in Fig. \ref{Figure5} is in qualitative agreement with the measured conductivity near superconducting transition in thin films (see, e.g. 
Refs. \cite{Armitage2011,Mondal2013,Nabeshima}).

\section{Conclusion}
\label{s5}

To summarize, we reported the general analytical expression for the quantum correction to the ac conductivity of a disordered electron system in the diffusive regime. In addition to the well established weak localization and Altshuler--Aronov corrections, we computed the contributions to the ac conductivity due to superconducting fluctuations above the transition temperature. 

In the static case, $\omega=0$, the weak localization, Altshuler--Aronov, and  DOS corrections can be 
resumed in the form of the one-loop terms in the renormalization group equation for the conductivity \cite{Fin1984Z}. The fluctuation propagator \eqref{eq:def:fl:prop} is also subjected to renormalization. In particular, the diffusion coefficient $D$ and dimensionless Cooper interaction $\gamma_c$ become scale dependent. Therefore, 
the contribution $\delta g_{f}^{CC}$ should be computed with the properly renormalized fluctuation propagator. For $\delta g_{f}^{CC}(\omega=0)$ such calculation results in the substitution of $\ln T/T_c$ by $1/\gamma_c(L_T)$. 
Here $L_T=\sqrt{D/T}$ stands for the length scale associated with the temperature (see Refs. \cite{BGM2015,BGM2016} for details). 

Present work can be extended in several ways. Our analysis can be extended to the pairing ac conductivity in the presence of a static magnetic field \cite{Galitski2001,KSF2012}. Also it would be tempting to study the effect of superconducting fluctuations on the physical observables in non-standard symmetry classes \cite{DellAnna2017}. The ac Nernst effect measured recently in thin superconducting films \cite{Frydman2020} suggests an interesting problem for computation of ac thermoelectric and thermal responses. 

\section{Acknowledgements}

The author is grateful to A. Levchenko, K. Tikhonov, A. Petkovi\'c, M. Skvortsov, and N. Stepanov for useful discussions. The author is indebted to I. Gornyi and A. Mirlin for fruitful collaboration at the initial stage of this work. The research was partially supported by the Russian Foundation for Basic Research (grant No. 20-52-12013) -- Deutsche Forschungsgemeinschaft (grant No. EV 30/14-1) cooperation and by the Alexander von Humboldt Foundation.

\appendix

\section{Anomalous Maki--Thompson contribution \label{App:MTan}} 

In this Appendix we present derivation for the asymptotic expression of the anomalous Maki--Thompson correction. Let us introduce the function
\begin{gather}
G(z) = \epsilon+\psi(z+1/2)-\psi(1/2) ,
\end{gather}
where $z=x+i y$. Then, we can rewrite Eq. \eqref{app:MT:an:t} as follows
\begin{gather}
\delta g^{\rm MT, an}(\omega) =  \frac{\sinh (2\pi \alpha)}{2\pi\alpha} 
\int\limits_0^\infty \frac{dx}{x-i\alpha+\gamma}
\int\limits_{-\infty}^\infty \frac{dy}{\sinh(2\pi y)}
\frac{1}{\sinh(2\pi(y+\alpha))} \frac{G(z)-G(z^*-2i\alpha)}{G(z)} .
\label{eq:App1:1}
\end{gather}
Here we introduced the following notations, $x=Dq^2/(4\pi T)$ and $y = \Omega/(4\pi T)$.

In the case $\alpha\gg 1$ it is convenient to rescale the integration variables as $x\to \alpha x$ and $y\to \alpha y$. Then, we find
\begin{gather}
\delta g^{\rm MT, an}(\omega) \approx  \frac{\sinh (2\pi \alpha)}{2\pi} 
\int\limits_0^\infty \frac{dx}{x-i}
\int\limits_{-\infty}^\infty \frac{dy}{\sinh(2\pi \alpha y)}
\frac{1}{\sinh(2\pi\alpha (y+1))} \frac{\ln [z/(z^*-2i)]}{\epsilon+\ln \alpha+\ln z-\psi(1/2)}
\notag \\
\approx - 
\int\limits_{-1}^0 \frac{dy}{\pi}    \int\limits_0^\infty \frac{dx}{x-i} \frac{\ln[{z}/({z^*-2i})]}{\epsilon+\ln \alpha+\ln z-\psi(1/2)}
\approx \frac{\pi^2-8\ln 2}{4\pi} \frac{1}{\epsilon+\ln \alpha}
- \frac{c^{\rm MT, an}}{(\epsilon+\ln \alpha)^2} ,
\end{gather}
where $c^{\rm MT, an}\approx 0.81 - 0.54 i$. We note that the main contribution to the integral comes from the region $x\sim y \sim \alpha \gg 1$.   

In the case of small frequencies, $\alpha\ll 1$, but away from the transition temperature, $\epsilon\gg 1$, we can expand Eq. \eqref{eq:App1:1} in $1/\epsilon$:
\begin{gather}
\delta g^{\rm MT, an}(\omega) = \frac{\alpha}{i \pi \epsilon} K_1+
\frac{1}{\epsilon^2} K_2 .
\label{eq:App1:I12}
\end{gather}
Here the first integral in the r.h.s. can be computed as follows
\begin{gather}
K_1 = \int\limits_0^\infty \frac{dx}{x-i\alpha+\gamma} \int\limits_{0}^\infty  dy \coth(2\pi y) \im \psi^{\prime\prime} (1/2+x+i y) 
= \int\limits_0^1 \frac{dx}{x-i\alpha+\gamma} \int\limits_{0}^\infty  dy
\coth(2\pi y) \frac{\pi^3\sinh(\pi y)}{\cosh^4(\pi y)}
\notag \\
+ \int\limits_0^1 \frac{dx}{x} \int\limits_{0}^\infty  dy 
\coth(2\pi y) \im \Bigl [ \psi^{\prime\prime} (1/2+x+i y)-\psi^{\prime\prime} (1/2+i y)\Bigr ]
+ \int\limits_1^\infty \frac{dx}{x} \int\limits_{0}^\infty  dy \coth(2\pi y)
\notag \\
\times
 \im \psi^{\prime\prime} (1/2+x+i y)
=\frac{2\pi^2}{3} \Bigl ( \ln \frac{1}{\gamma-i \alpha} - c^{\rm MT, an}_1\Bigr ) ,
\end{gather}
where the numerical constant is equal $c_1\approx 1.62$. The second integral in the r.h.s. of Eq. \eqref{eq:App1:I12} can be evaluated as 
\begin{gather}
K_2=4  \int\limits_0^\infty \frac{dx}{x-i\alpha+\gamma} \int\limits_{0}^\infty \frac{dy \bigl[\im \psi(1/2+x+i y)\bigr ]^2}{\sinh^2(2\pi y)} 
= 4 \Biggl\{  \frac{\pi^2}{4} \int\limits_0^1 \frac{dx}{x-i\alpha+\gamma}
\int\limits_{0}^\infty \frac{dy \tanh^2(\pi y)}{\sinh^2(2\pi y)}
\notag \\
+ \int\limits_1^\infty \frac{dx}{x} \int\limits_{0}^\infty \frac{dy \bigl[\im \psi(1/2+x+i y)\bigr ]^2}{\sinh^2(2\pi y)}
+ \int\limits_0^1 \frac{dx}{x} \int\limits_{0}^\infty \frac{dy}{\sinh^2(2\pi y)}
\Bigl[ \bigl[\im \psi(1/2+x+i y)\bigr ]^2 
\notag \\
- \bigl[\im \psi(1/2+i y)\bigr ]^2\Bigr ] 
\Biggr \}
=\frac{\pi}{6} \Bigl ( \ln \frac{1}{\gamma-i \alpha} -c^{\rm MT, an}_2\Bigr ) .
\end{gather}
Here the numerical constant is equal $c^{\rm MT, an}_2\approx 2.19$.
Finally,
\begin{gather}
\delta g^{\rm MT, an}(\omega) = \frac{2\pi \alpha}{3i \epsilon}\Bigl ( \ln \frac{1}{\gamma-i\alpha}-c^{\rm MT, an}_1\Bigr ) + \frac{\pi}{6\epsilon^2} \Bigl ( \ln \frac{1}{\gamma-i \alpha} -c^{\rm MT, an}_2\Bigr ) .
\end{gather}

Finally, we consider the region $\epsilon\ll 1$ and $\alpha\ll 1$. Then we can split the anomalous Maki-Thompson correction into four parts
\begin{gather}
\delta g^{\rm MT, an}(\omega) \approx
I_1+I_2 \ln (\gamma-i\alpha) +I_3+I_4 .
\end{gather}
The first contribution can be estimated as follows
\begin{gather}
I_1=\int\limits_0^1 \frac{dx}{x-i\alpha+\gamma}
\int\limits_{-1}^1 \frac{dy}{\sinh(2\pi y)\sinh(2\pi (y+\alpha))}
\frac{\psi(1/2+x+i y)-\psi(1/2+x-i y-2i\alpha)}{\epsilon+\psi(1/2+x+i y)-\psi(1/2)}
\notag \\
\approx \frac{i}{2\pi^2}\int\limits_0^1 \frac{dx}{(x-i\alpha+\gamma)}
\int\limits_{-\infty}^\infty \frac{dy}{y}
\frac{1}{\bar{\epsilon}+x+i y} = -\frac{1}{2\pi} \frac{1}{\bar\epsilon-\gamma+i\alpha}\ln 
\frac{\bar{\epsilon}}{\gamma-i\alpha},
\end{gather}
where $\bar\epsilon = 2\epsilon/\pi^2$. We note that there are also subleading terms proportional to $\ln \epsilon$. The other three contributions can be approximated by their values at $\epsilon=\alpha=0$,
\begin{gather}
I_2=- 4
\int\limits_{1}^\infty \frac{dy}{\sinh^2(2\pi y)}
\left |\frac{\im \psi(1/2+i y)}{\psi(1/2+i y)-\psi(1/2)}\right |^2
\approx - 1.7 \cdot 10^{-6} ,
\end{gather}
\begin{gather}
I_3 = 4  \int\limits_0^1 \frac{dx}{x}
\int\limits_{1}^\infty \frac{dy}{\sinh^2(2\pi y)}
\Biggl [ \left |\frac{\im \psi(1/2+x+i y)}{\psi(1/2+x+i y)-\psi(1/2)}\right |^2
- \left |\frac{\im \psi(1/2+i y)}{\psi(1/2+i y)-\psi(1/2)}\right |^2
\Biggr ] \notag \\
\approx -1.4\cdot  10^{-6} ,
\end{gather}
and
\begin{gather}
\notag \\
I_4=4  \int\limits_1^\infty \frac{dx}{x}\int\limits_{0}^\infty \frac{dy}{\sinh^2(2\pi y)}\left |\frac{\im \psi(1/2+x+i y)}{\psi(1/2+x+i y)-\psi(1/2)}\right |^2
\approx 0.0021 .
\end{gather}

\section{DOS correction\label{App_DOS}}

In this Appendix we present derivation for the asymptotic expression of the DOS correction. We start from splitting the expression \eqref{eq_Sigma_DOS} into two parts
\begin{equation}
\delta g^{\rm DOS}(\omega) = \delta g^{\rm DOS}_1(\omega) +
\delta g^{\rm DOS}_2(\omega) ,
\end{equation}
where
\begin{gather}
\delta g^{\rm DOS}_1(\omega) =
\int\limits_0^\infty \frac{dx}{4\pi \alpha}\int\limits_{-\infty}^\infty dy 
 \bigl [\coth(2\pi(y-\alpha)-\coth(2\pi y)\bigr ]
\frac{G^\prime(z^*)- 
G^\prime(z-2i \alpha)}{G(z^*)}  ,
\notag \\
\delta g^{\rm DOS}_2(\omega) =
-\int\limits_0^\Lambda \frac{dx}{4\pi \alpha}\int\limits_{-\Lambda}^\Lambda dy \coth(2\pi y) 
\frac{G^\prime(z^*)- 
G^\prime(z^*-2i \alpha)}{G(z^*)} .
\end{gather}
Here we introduced the dimensionless ultra-violet cut off $\Lambda=1/(4\pi T\tau) \gg 1$. 
Next we split $\delta g^{\rm DOS}_2(\omega)$ into three terms
\begin{gather}
\delta g^{\rm DOS}_2(\omega) =
\delta g^{\rm DOS}_{2,1}(\omega) +
\delta g^{\rm DOS}_{2,2}(\omega)+
\delta g^{\rm DOS}_{2,3}(\omega) .
\end{gather}
The first two term is organized in such a way that one can integrate over $x$ exactly,
\begin{gather}
\delta g^{\rm DOS}_{2,1}(\omega) =
- \int\limits_0^\Lambda \frac{dy}{4\pi \alpha}
\ln \frac{G(-i y-2i\alpha)G(i y)}{G(-i y)G(i y-2i\alpha)}
- \int\limits_0^{2\alpha} \frac{dy}{2\pi \alpha} \ln G(-i y) .
\end{gather}
The other two contributions are given as
\begin{gather}
\notag \\
\delta g^{\rm DOS}_{2,2}(\omega) = \int\limits_0^{2\alpha} \frac{dy}{2\pi \alpha} \ln G(-i y)
+
 \int\limits_0^\infty \frac{dx}{4\pi \alpha}\int\limits_{0}^\infty dy
\bigl [1-\coth(2\pi y)\bigr ] \Bigl [ \frac{G^\prime(z^*)-G^\prime(z^*-2i\alpha)}{G(z^*)} 
\notag \\
-  \frac{G^\prime(z)-G^\prime(z-2i\alpha)}{G(z)}
\Bigr ] ,
\label{eq:App2:2a}
\end{gather}
and
\begin{gather}
\delta g^{\rm DOS}_{2,3}(\omega) = 
-\int\limits_0^\infty \frac{dx}{4\pi \alpha}\int\limits_{0}^\infty dy
\left [ \frac{G^\prime(z^*-2i\alpha)}{G(z^*-2i\alpha)}-
\frac{G^\prime(z^*-2i\alpha)}{G(z^*)}
-
\frac{G^\prime(z-2i\alpha)}{G(z-2i\alpha)}+
\frac{G^\prime(z-2i\alpha)}{G(z)}\right ] 
 .\label{eq:App2:3}
\end{gather}

The integral over $y$ in the expression for $\delta g^{\rm DOS}_{2,1}(\omega)$ can be performed exactly,
\begin{gather}
\delta g^{\rm DOS}_{2,1}(\omega)  = 
- \int\limits_\Lambda^{\Lambda+2\alpha} \frac{dy}{4\pi \alpha}
\ln G(-i y) 
- \int\limits_{\Lambda-2\alpha}^{\Lambda} \frac{dy}{4\pi \alpha}
\ln G(-i y)
= -\frac{1}{\pi} \ln G(-i \Lambda)= - \frac{1}{\pi}\ln (\epsilon+\ln \Lambda) .
\end{gather}
Then we obtain Eq. \eqref{eq:DOS:split} in which
$\delta g^{\rm DOS}_{f}(\omega) = \delta g^{\rm DOS}_{1}(\omega) +\delta g^{\rm DOS}_{2,2}(\omega)+\delta g^{\rm DOS}_{2,3}(\omega)$.

In the case of large frequencies, $\alpha \gg 1$, it is convenient to perform rescaling $x \to \alpha x$ and $y \to \alpha y$. Then we obtain
\begin{gather}
\delta g^{\rm DOS}_1(\omega) =\frac{1}{\epsilon+\ln\alpha}
\int\limits_0^\infty \frac{dx}{4\pi}\int\limits_{0}^1 dy
 \bigl [ \ln y - \ln(2-y)\bigr ]
 =-\frac{\ln 2}{\pi} \frac{1}{\epsilon+\ln\alpha}.
\end{gather}
Neglecting the second integral in the right hand side of Eq. \eqref{eq:App2:2a}, we find in a similar way 
\begin{equation}
\delta g^{\rm DOS}_{2,2}(\omega) =
\frac{1}{\pi} \ln (\epsilon+\ln \alpha) - \frac{i}{2}\frac{1}{\epsilon+\ln \alpha} .
\end{equation}
Next, we find
\begin{gather}
\delta g^{\rm DOS}_{2,3}(\omega) =
-\int\limits_0^\infty \frac{dx}{4\pi}\int\limits_{0}^\infty dy
\Biggl [ \frac{1}{z^*-2i} \frac{\ln[{z^*}/({z^*-2i})]}{(\epsilon+\ln\alpha+\ln z^*)(\epsilon+\ln\alpha+\ln(z^*-2i))}
\notag \\ -
\frac{1}{z-2i} \frac{\ln[{z}/({z-2i})]}{(\epsilon+\ln\alpha+\ln z)(\epsilon+\ln\alpha+\ln(z-2i))} 
\Biggr ] 
\notag \\
\approx -\frac{1}{\pi} \int\limits_{\sim 1}^\infty dx\int\limits_{\sim 1}^\infty dy
\im \left [\frac{1}{z(\epsilon+\ln\alpha+\ln z)}\right]^2
\approx \frac{1}{\pi} \int\limits_{\sim 1}^\infty \frac{dr}{r} \frac{1}{(\epsilon+\ln\alpha+\ln r)^2}
=\frac{1}{\pi} \frac{1}{\epsilon+\ln\alpha}
\end{gather}

In the case $\alpha\ll 1$ and $\epsilon\gg 1$, we expand the integrand in $\delta g^{\rm DOS}_1(\omega)$ in series in $1/\epsilon$ and obtain
\begin{gather}
\delta g^{\rm DOS}_1(\omega) = - \frac{i}{2\epsilon^2}
\int\limits_{-\infty}^\infty \frac{dy}{\sinh^2(2\pi y)} 
\int\limits_0^\infty dx\, \partial_x \Bigl [ \psi(1/2+x-i y) 
\im \psi(1/2+x-i y)\Bigr ]
=-\frac{\pi}{24 \epsilon^2} .
\end{gather}
In a similar way, we find
\begin{gather}
\delta g^{\rm DOS}_{2,2}(\omega) =
\frac{1}{\pi} \ln \epsilon - \frac{1}{\pi \epsilon}\int\limits_0^\infty dy [1-\coth(2\pi y)]
\im \psi^\prime(1/2+i y) 
-\frac{\pi i \alpha}{2\epsilon} +\frac{\alpha i}{\pi \epsilon}
\int\limits_0^\infty dy [1-\coth(2\pi y)]
\notag \\
\times 
\im \psi^{\prime\prime}(1/2+i y) 
= \frac{1}{\pi} \ln \epsilon + \frac{\ln 2-1}{\pi \epsilon}- \frac{2 \pi i \alpha}{3\epsilon} ,
\end{gather}
Next, we can write
\begin{gather}
\delta g^{\rm DOS}_{2,3}(\omega) \approx 
-\frac{1}{\pi} \int\limits_0^\infty dx \int\limits_{0}^\infty dy
\im \left [ \frac{G^\prime(z)}{G(z)}\right ]^2\approx  \frac{1}{\pi} \int\limits_{\sim 1}^\infty \frac{dr}{r}
\frac{1}{(\epsilon+\ln r)^2}
= \frac{1}{\pi \epsilon} ,
\end{gather}

Finally, we consider small frequencies, $\alpha\ll 1$, and temperatures close to the superconducting transition, $\epsilon\ll 1$. At first, we split $\delta g^{\rm DOS}_1(\omega)$ into three parts 
\begin{gather}
\delta g^{\rm DOS}_1(\omega) = 
\frac{\sinh(2\pi \alpha)}{4\pi \alpha} \int\limits_0^1 dx \int\limits_{-1}^1  
\frac{dy \, [G^\prime(z^*)- 
G^\prime(z-2i \alpha)]}{\sinh(2\pi(y-\alpha))\sinh(2\pi y)G(z^*)}
+2  \int\limits_1^\infty dx \int\limits_{0}^1  
\frac{dy}{\sinh^2(2\pi y)}\notag \\
\times
 \frac{\im 
\psi^\prime(1/2+z) \im 
\psi(1/2+z)}{\bigl |\psi(1/2+z)-\psi(1/2)\bigr |^2}
+
2\int\limits_0^\infty dx \int\limits_{1}^\infty \frac{dy}{\sinh^2(2\pi y)}
\frac{\im 
\psi^\prime(1/2+z) \im 
\psi(1/2+z)}{\bigl |\psi(1/2+z)-\psi(1/2)\bigr |^2}.
\label{eq:App2:4}
\end{gather}
Here we neglected $\alpha$ and $\epsilon$ whenever it is possible. Next, we omit the terms independent of $\alpha$ and $\epsilon$ and expand the integrand in the first line of Eq. \eqref{eq:App2:4} to the lowest order in $x$, $y$, and $\alpha$. Then we find with the logarithmic accuracy, 
\begin{gather}
\delta g^{\rm DOS}_1(\omega) = 
\frac{1}{4\pi^2} \int\limits_0^1 dx \int\limits_{-1}^1  
\frac{dy}{y}
\frac{i \psi^{\prime\prime}(1/2)+\alpha\psi^{\prime\prime\prime}(1/2)}{\epsilon+\psi^\prime(1/2)(x-i y)}
= -\left (\frac{7 \zeta(3)}{\pi^3}+ \frac{i\pi\alpha}{2}\right )\ln \frac{1}{\epsilon} .
\label{eq:App2:5}
\end{gather}
Next we find
\begin{gather}
\delta g^{\rm DOS}_{2,2}(\omega) = \int\limits_0^{2\alpha} \frac{dy}{2\pi\alpha} \ln [\epsilon-i  \psi^\prime(1/2) y]+
\int\limits_0^1 \frac{dx}{4\pi \alpha}\int\limits_{0}^\infty dy
\bigl [1-\coth(2\pi y)\bigr ] \Bigl [ \frac{G^\prime(z^*)-G^\prime(z^*-2i\alpha)}{G(z^*)} \notag \\
 -  \frac{G^\prime(z)-G^\prime(z-2i\alpha)}{G(z)}
\Bigr ] +\int\limits_1^\infty \frac{dx}{\pi}\int\limits_{0}^\infty dy
\bigl [1-\coth(2\pi y)\bigr ] \im  \frac{G^{\prime\prime}(z)}{G(z)} 
\notag \\
\approx \frac{1}{\pi} \ln \epsilon+ \frac{1}{\pi}
\left (1+\frac{i \epsilon}{\pi^2\alpha} \right )\ln \left (1-\frac{i \pi^2\alpha}{\epsilon}\right ) +
\frac{\psi^{\prime\prime}(1/2)- i \pi^4\alpha}{2\pi \psi^\prime(1/2)}
\int\limits_0^1 dx \int\limits_{0}^\infty \frac{dy}{(\bar\epsilon+x)^2+y^2}
\end{gather}
Hence, we find with the logarithmic accuracy
\begin{gather}
\delta g^{\rm DOS}_{2,2}(\omega) = -\left (\frac{7 \zeta(3)}{\pi^3}+ \frac{1}{\pi}+\frac{i\pi\alpha}{2}\right ) \ln \frac{1}{\epsilon}+ \frac{1}{\pi}
\left (1+\frac{i \epsilon}{\pi^2\alpha} \right )\ln \left (1-\frac{i \pi^2\alpha}{\epsilon}\right )
.
\end{gather} 
Also, we obtain with logarithmic accuracy
\begin{gather}
\delta g^{\rm DOS}_{2,3}(\omega) = 
-\frac{1}{\pi} \int\limits_0^\infty dx \int\limits_{0}^\infty dy
\im \left [ \frac{G^\prime(z)}{G(z)}\right ]^2 =
-\frac{1}{\pi} \int\limits_{0}^1 dy\int\limits_0^\infty dx  
\im \left [ \frac{1}{\bar\epsilon+x+ i y}\right ]^2
\notag \\
-\frac{1}{\pi} \int\limits_0^\infty dx \int\limits_{1}^\infty dy
\im \left [ \frac{G^\prime(z)}{G(z)}\right ]^2
= \frac{1}{\pi} \int\limits_{0}^1 dy \frac{y}{\bar\epsilon^2+ y^2}
-\frac{1}{\pi} \int\limits_0^\infty dx \int\limits_{1}^\infty dy
\im \left [ \frac{G^\prime(z)}{G(z)}\right ]^2 
= \frac{1}{\pi}\ln \frac{1}{\epsilon} .
\end{gather}

\section{Aslamazov--Larkin contribution\label{App:AL:Sec}}

In this Appendix we present derivation for the asymptotic expression of the Aslamazov--Larkin contribution. This correction can be written as 
\begin{gather}
\delta g^{\rm AL}(\omega) = 
-\frac{\sinh(2\pi \alpha)}{2\pi \alpha^3} \int\limits_0^\infty dx\ x\int\limits_{-\infty}^\infty \frac{dy}{\sinh(2\pi y)\sinh \bigl (2\pi(y-\alpha)\bigr )}
\frac{\im G(z-i\alpha)}{|G(z-i\alpha)|^2} \frac{G(z^*+i\alpha)- G(z^*-i\alpha)}{G(z^*)} 
\notag \\
\times
\im \Bigr[ G(z^*+i\alpha)- G(z^*-i\alpha)\Bigr ] .
\label{eq:App:AL}
\end{gather}

In the case of large frequencies, $\alpha \gg 1$, it is convenient to perform rescaling $x \to \alpha x$ and $y \to \alpha y$. Then we obtain
\begin{gather}
\delta g^{\rm AL}(\omega) = 
\frac{1}{\pi}\frac{1}{(\epsilon+\ln\alpha)^3} \int\limits_0^\infty dx\ x\int\limits_{0}^1 dy 
\arctan \left (\frac{y-1}{x}\right )
\ln \frac{x-i y+i}{x-i y-i}\,
\Biggl [\arctan \left (\frac{1-y}{x}\right )
\notag \\
+\arctan \left (\frac{1+y}{x}\right )\Biggr ]
\approx \frac{c^{\rm AL}_3}{(\epsilon+\ln\alpha)^3} ,
\end{gather}
where the constant $c^{\rm AL}_3 \approx 0.17 - 0.89 i$.

In the case of small frequencies, $\alpha\ll 1$, and high temperatures, $\epsilon\gg 1$, we can approximate the function $G(z)$ in denominators of the integrand in Eq. \eqref{eq:App:AL} by $\epsilon$,  
\begin{gather}
\delta g^{\rm AL}(\omega) \approx
\frac{4i}{\epsilon^3} \int\limits_0^\infty dx\ x\int\limits_{-\infty}^\infty dy 
\ \partial_x f(x,y) \im f(x,y-\alpha)\re \psi^\prime(1/2+x-iy) ,\label{eq:App:AL2}
\end{gather}
where $f(x,y) = \psi(1/2+x-iy)/\sinh(2\pi y)$. 
Expanding in $\alpha$ in the right hand side of Eq. \eqref{eq:App:AL2}, we obtain
\begin{gather}
\delta g^{\rm AL}(\omega) \approx \frac{c^{\rm AL}_4-c^{\rm AL}_5 i \alpha}{\epsilon^3} ,
\end{gather}
where $c^{\rm AL}_4\approx 1.44$ and $c^{\rm AL}_5\approx 9.23$. 

In the vicinity of the superconducting transition, $\epsilon\ll 1$, and for small frequencies, $\alpha\ll 1$, we can expand the integrand in Eq. \eqref{eq:App:AL} in $y$ and $x$,
\begin{gather}
\delta g^{\rm AL}(\omega) \approx -\frac{i}{\pi^2}  \int\limits_0^\infty dx\int\limits_{-\infty}^\infty \frac{dy}{y} \frac{x}{[(\bar\epsilon+x)^2+(y-\alpha)^2][\bar\epsilon+x-iy]}
= \frac{\pi}{8 \epsilon} W_1\left (\frac{\pi^2\alpha}{2\epsilon}\right ) 
- \frac{i \pi^3\alpha}{32\epsilon^2}  W_2\left (\frac{\pi^2\alpha}{2\epsilon}\right )  ,
\end{gather}
where the functions $W_1(X)$ and $W_2(X)$ are defined in Eq. \eqref{eq:W12}.
We note that there are also subleading terms proportional to $\ln \epsilon$.

\section{The correction $\delta g^{\rm sc}(\omega)$ \label{App:gsc:Sec}}

In this Appendix we present derivation for the asymptotic expression of the correction $\delta g^{\rm sc}(\omega)$. It is convenient to split the expression \eqref{eq:sigma:sc:f}  into four parts, $\delta g^{\rm sc}(\omega)= \delta g^{\rm sc}_{I}(\omega)+\delta g^{\rm sc}_{II}(\omega)+\delta g^{\rm sc}_{III}(\omega)+\delta g^{\rm sc}_{IV}(\omega)$, and discuss each of them separately.

\subsection{$\delta g^{\rm sc}_I(\omega)$}

The first contribution $\delta g^{\rm sc}_I(\omega)$ can be expressed in terms of the dimensionless parameters in the following way
\begin{gather}
\delta g^{\rm sc}_I(\omega) = 
\frac{1}{4\pi\alpha} \int\limits_{0}^\infty dx\ x \int \limits_{-\infty}^\infty \frac{dy}{G(z)} \coth(2\pi y) \Biggl \{ 3G^{\prime\prime}(z)+G^{\prime\prime}(z-2i\alpha)
+\frac{2}{\alpha^2} \Bigl[ G(z-2i\alpha)-G(z-i\alpha)
\notag \\
-G(z)+G(z+i\alpha)\Bigr]\Biggr \} .
\label{eq:app:D1}
\end{gather}

In the case of large frequencies, $\alpha\gg 1$, we perform rescaling $x\to \alpha x$ and $y\to \alpha y$. Then we find
\begin{gather}
\delta g^{\rm sc}_I(\omega) \approx \frac{1}{4\pi}\frac{1}{\epsilon+\ln\alpha}
\int \limits_0^\infty dx\ x \int \limits_{-\infty}^\infty dy \sgn y
\Biggl [
-\frac{3}{z^2}-\frac{1}{(z-2i)^2} 
+ 2\ln \frac{(z-2i)(z+i)}{(z-i)z}\Biggr ]
\notag \\
= \frac{1}{6\pi} \frac{5-2\ln 2 + i\pi}{\epsilon+\ln\alpha} .
\end{gather}
At low frequencies, $\alpha\ll 1$, and at high temperatures, $\epsilon\gg 1$, we expand Eq. \eqref{eq:app:D1} in $\alpha$ and obtain 
\begin{gather}
\delta g^{\rm sc}_I(\omega) \approx 
-\frac{\alpha }{6\pi\epsilon} \int\limits_{0}^\infty dx\ x \int \limits_{-\infty}^\infty dy \coth(2\pi y)\ G^{\prime\prime\prime\prime}(z) 
+\frac{5 i\alpha^2 }{24\pi\epsilon} \int\limits_{0}^\infty dx\ x \int \limits_{-\infty}^\infty dy \coth(2\pi y)\ G^{\prime\prime\prime\prime\prime}(z) 
\notag \\
= -\frac{2\pi i\alpha}{9\epsilon}+\frac{c_6\alpha^2}{\epsilon},
\end{gather}
where $c_6\approx 3.83$.

Near the superconducting transition, $\epsilon\ll 1$, and for small frequencies, $\alpha\ll 1$,  the correction $\delta g^{\rm sc}_I(\omega)$ does not diverge in the limit $\epsilon\to 0$.

\subsection{$\delta g^{\rm sc}_{II}(\omega)$}

The contribution $\delta g^{\rm sc}_{II}(\omega)$ reads
\begin{gather}
\delta g^{\rm sc}_{II}(\omega) = 
\frac{1}{4\pi\alpha} \int\limits_{0}^\infty dx\ x \int \limits_{-\infty}^\infty \frac{dy}{G(z)} \Bigl[\coth\bigl (2\pi (y-\alpha)\bigr )- \coth(2\pi y)\Bigr] \Biggl \{ G^{\prime\prime}(z-i\alpha)-G^{\prime\prime}(z^*)
\notag\\
+\frac{2i}{\alpha} \Bigl [ G^\prime(z^*) +\frac{G(z^*)-G(z^*+i\alpha)}{i\alpha}
-G^\prime(z-2i\alpha) - \frac{G(z-2i\alpha)-G(z-i\alpha)}{i\alpha}
\Bigr]\Biggr \} .
\end{gather}
In the case of large frequencies, $\alpha\gg 1$, it is convenient to rescale integration variables $x\to \alpha x$ and $y\to \alpha y$. Hence, we obtain
\begin{gather}
\delta g^{\rm sc}_{II}(\omega) \approx 
-\frac{1}{2\pi}\frac{1}{\epsilon+\ln\alpha} \int\limits_{0}^\infty dx\ x \int \limits_{0}^1 dy \Biggl \{ -\frac{1}{(z-i)^2}+\frac{1}{z^{*2}}
+ \frac{2i}{z^*}-\frac{2i}{z-2i} +2 \ln\frac{z^*(z-i)}{(z^*+i)(z-2i)}
\Biggr \}\notag\\
=\frac{1}{6\pi}\frac{14\ln 2-4-i\pi}{\epsilon+\ln\alpha}  .
\end{gather}
In the case of small frequencies, $\alpha\ll 1$, but well above the superconductivity transition temperature, $\epsilon \gg 1$, we expand $\delta g^{\rm sc}_{II}(\omega)$ in $\alpha$. Then we find 
\begin{gather}
\delta g^{\rm sc}_{II}(\omega) \approx 
\frac{\sinh(2\pi \alpha)}{4\pi\alpha \epsilon}  \int \limits_{-\infty}^\infty \frac{dy}{\sinh\bigl (2\pi (y-\alpha)\bigr )\sinh(2\pi y)} 
\Biggl \{ \frac{i\alpha}{3} \Bigl [\psi^\prime\left({1}/{2}-i y\right ) + 2 \psi^\prime\left({1}/{2}+i y\right )\Bigr ]
\notag \\
 - \frac{\alpha^2}{12}\Bigl [\psi^{\prime\prime}\left({1}/{2}-i y\right ) -11  \psi^{\prime\prime}\left({1}/{2}+i y\right )\Bigr ]
 \Biggr \} = -\frac{i\pi\alpha}{3\epsilon}+ \frac{c_6\alpha^2}{\epsilon} .
\end{gather}

The correction $\delta g^{\rm sc}_{II}(\omega)$ becomes a constant in the limit 
$\alpha\ll 1$ and $\epsilon\ll 1$.

\subsection{$\delta g^{\rm sc}_{III}(\omega)$ and $\delta g^{\rm sc}_{IV}(\omega)$}

The contributions $\delta g^{\rm sc}_{III}(\omega)$ and $\delta g^{\rm sc}_{IV}(\omega)$ are given as
\begin{gather}
\delta g^{\rm sc}_{III}(\omega) = 
- \frac{1}{4\pi\alpha} \int\limits_{0}^\infty dx\ x \int \limits_{-\infty}^\infty \frac{dy}{G(z)} \coth(2\pi y)\Biggl \{ \frac{G^\prime(z)}{G(z)}\Biggl[3G^\prime(z)+G^\prime(z-2i\alpha)
+2 \frac{G(z)-G(z-2i\alpha)}{i\alpha}\Biggr ]
\notag \\
+ 2 \frac{\bigl [G(z+i\alpha)-G(z-i\alpha)\bigr]^2}{\alpha^2G(z+i\alpha)}
\Biggr \} 
\end{gather}
and
\begin{gather}
\delta g^{\rm sc}_{IV}(\omega) = 
\frac{1}{4\pi\alpha} \int\limits_{0}^\infty dx\ x \int \limits_{-\infty}^\infty \frac{dy}{G(z^*)} \Bigl[\coth\bigl (2\pi (y-\alpha)\bigr )- \coth(2\pi y)\Bigr] 
\Biggl \{ \frac{G^\prime(z^*)}{G(z^*)} 
\Bigl [ G^{\prime}(z^*)-G^{\prime}(z-2i\alpha)\Bigr ]
\notag\\
+\frac{G(z^*+i\alpha)-G(z^*-i\alpha)}{\alpha^2 G(z^*+i\alpha)}
\Bigl [G(z^*+i\alpha)-G(z^*-i\alpha)+ \re G(z^*+i\alpha)-\re G(z^*-i\alpha)\Bigr ]
\notag \\
-
 \frac{G(z^*+i\alpha)-G(z^*-i\alpha)}{i\alpha^2 G(z-i\alpha)} \im \Bigl[ G(z^*+i\alpha)-G(z^*-i\alpha)\Bigr ]
\notag \\
- \frac{\bigl [G(z-i\alpha)-G(z-2i\alpha)+G(z^*)-G(z^*-i\alpha)\bigr ]^2}{\alpha^2 G(z-i\alpha)}
\Biggr \} .
\end{gather}

It is convenient to split the contribution $\delta g^{\rm sc}_{IV}(\omega)$ into two parts. The first part is given as
\begin{gather}
\delta g^{\rm sc}_{IV,1}(\omega) = 
\frac{\sinh(2\pi\alpha)}{4\pi\alpha} \int\limits_{0}^\infty dx\ x \int \limits_{-\infty}^\infty dy \frac{G^\prime(z^*)}{G^2(z^*)} \frac{G^{\prime}(z^*)-G^{\prime}(z-2i\alpha)}{\sinh\bigl (2\pi (y-\alpha)\bigr )\sinh(2\pi y)} .
\end{gather}
The second part of $\delta g^{\rm sc}_{IV}(\omega)$ can be combined with the term $\delta g^{\rm sc}_{III}(\omega)$. Then we find
\begin{gather}
\delta g^{\rm sc}_{III}(\omega) + \delta g^{\rm sc}_{IV,2}(\omega) =
-\frac{1}{4\pi\alpha} \int\limits_{0}^\infty dx\ x \int \limits_{-\infty}^\infty dy \coth(2\pi y)
\Biggl\{
\frac{G^\prime(z)}{G^2(z)}\Biggl[3G^\prime(z)+G^\prime(z-2i\alpha)
\notag \\
+2 \frac{G(z)-G(z-2i\alpha)}{i\alpha}\Biggr ]
+ 2 \frac{\bigl [G(z+i\alpha)-G(z-i\alpha)\bigr]^2}{\alpha^2G(z) G(z+i\alpha)}
+
\frac{G(z^*+i\alpha)-G(z^*-i\alpha)}{\alpha^2 G(z^*)G(z^*+i\alpha)}
\notag \\
\times
\Bigl [G(z^*+i\alpha)-G(z^*-i\alpha)+ \re G(z^*+i\alpha)-\re G(z^*-i\alpha)\Bigr ]
\notag \\
-
 \frac{G(z^*+i\alpha)-G(z^*-i\alpha)}{i\alpha^2 G(z^*)G(z-i\alpha)} \im \Bigl[ G(z^*+i\alpha)-G(z^*-i\alpha)\Bigr ]
\notag \\
- \frac{\bigl [G(z-i\alpha)-G(z-2i\alpha)+G(z^*)-G(z^*-i\alpha)\bigr ]^2}{\alpha^2 G(z^*)G(z-i\alpha)}
-
\frac{G(z^*)-G(z^*-2i\alpha)}{\alpha^2 G(z^*-i\alpha)G(z^*)}
\notag \\
\times
\Bigl [G(z^*)-G(z^*-2i\alpha)+ \re G(z^*)-\re G(z^*-2i\alpha)\Bigr ]
\notag \\
+
 \frac{G(z^*)-G(z^*-2i\alpha)}{i\alpha^2 G(z^*-i\alpha)G(z)} \im \Bigl[ G(z^*)-G(z^*-2i\alpha)\Bigr ]
\notag \\
+ \frac{\bigl [G(z)-G(z-i\alpha)+G(z^*-i\alpha)-G(z^*-2i\alpha)\bigr ]^2}{\alpha^2 G(z^*-i\alpha)G(z)}
\Biggr \} .
\label{eq:III+IV2}
\end{gather}

At first, we consider the regime of large frequencies, $\alpha\gg 1$. It is convenient to make the rescaling $x\to \alpha x$ and $y\to \alpha y$. Then we obtain
\begin{gather}
\delta g^{\rm sc}_{IV,1}(\omega) \approx 
\frac{i}{\pi(\epsilon+\ln\alpha)^2} \int\limits_{0}^\infty dx\ x \int \limits_{0}^1 dy
\frac{1}{z^{*2}(z^*-2i)}= 
-\frac{1}{8\pi} \frac{9\ln 3-4\ln 2-2}{(\epsilon+\ln\alpha)^2} .
\end{gather}
Also, we find

\begin{gather}
\delta g^{\rm sc}_{III}(\omega) +\delta g^{\rm sc}_{IV,2}(\omega) \approx 
- \frac{1}{4\pi} \int\limits_{0}^\infty dx\ x \int \limits_{-\infty}^\infty dy
\frac{\sgn y}{(\epsilon+\ln\alpha+\ln|z|)^2} \Biggl\{
\frac{1}{z}\Bigl [ \frac{3}{z}+\frac{1}{z-2i}-2i \ln\frac{z}{z-2i} \Bigr ]
\notag \\
+2\ln^2\frac{z+i}{z-i}
+2 \ln^2 \frac{z^*+i}{z^*-i}
-\ln^2 \frac{(z-i)z^*}{(z^*-i)(z-2i)}
-2 \ln^2 \frac{z^*}{z^*-2i}
+\ln^2 \frac{z (z^*-i)}{(z^*-2i)(z-i)}
\Biggr ] 
\notag \\
\approx \frac{1}{2\pi} \int\limits_{\sim 1}^\infty \frac{dr}{r}
\frac{1}{(\epsilon+\ln\alpha+\ln r)^2} = 
\frac{1}{2\pi} \frac{1}{\epsilon+\ln\alpha} .
\end{gather}

Next, we consider the case of small frequencies, $\alpha\ll 1$, and high temperatures, $\epsilon\gg 1$. Then, expanding in $\alpha$, we find
\begin{gather}
\delta g^{\rm sc}_{IV,1}(\omega) \approx 
-\frac{2}{\epsilon^2} \int\limits_{0}^\infty dx\ x \int \limits_{-\infty}^\infty dy \frac{[\im G^\prime(z)]^2}{\sinh^2(2\pi y)} \approx - \frac{c_7}{\epsilon^2} ,
\end{gather}
where $c_7\approx 0.047$. For the contribution $\delta g^{\rm sc}_{III}(\omega) +\delta g^{\rm sc}_{IV,2}(\omega)$  
the integrals over $x$ and $y$ are dominated by their large values of the order of $\exp \epsilon$. Therefore, after expansion in $\alpha$, we obtain
\begin{gather}
\delta g^{\rm sc}_{III}(\omega) +\delta g^{\rm sc}_{IV,2}(\omega) \approx 
\frac{1}{\pi}
\int\limits_{0}^\infty dx\ x \int \limits_{0}^\infty dy \frac{\coth(2\pi y)}{(\epsilon+\ln|z|)^2}\im 
G^\prime(z) \Bigl [3G^{\prime\prime}(z) +2G^{\prime\prime}(z^*)\Bigr ] 
\notag \\
\approx \frac{1}{2\pi}\int\limits_{\sim 1}^\infty\frac{dr}{r} \frac{1}{(\epsilon+\ln r)^2}
= \frac{1}{2\pi \epsilon}.
\end{gather}
We note that the terms of the next order in $\alpha$ has additional smallness in $1/\epsilon$.

Finally, we consider the vicinity of the superconducting transition, $\epsilon\ll 1$, and small frequencies, $\alpha\ll 1$. Then expanding in $x$ and $y$ we find
\begin{gather}
\delta g^{\rm sc}_{IV,1}(\omega) \approx 
\frac{\psi^{\prime\prime}(1/2)}{\pi^2\psi^\prime(1/2)} \int\limits_{0}^{\sim 1} dx \int \limits_{0}^\infty dy \frac{x(\bar\epsilon+x)}{[(\bar\epsilon+x)^2+y^2]^2} 
= \frac{7\zeta(3)}{\pi^3}\ln \epsilon
.
\end{gather}
Here we neglected the dependence on $\alpha$ since it does not lead to terms divergent for $\epsilon \to 0$. In order to analyse the term $\delta g^{\rm sc}_{III}(\omega) +\delta g^{\rm sc}_{IV,2}(\omega)$, at first,  we perform expansion of enumerators in $\alpha$ in the right hand side of Eq. \eqref{eq:III+IV2},
\begin{gather}
\delta g^{\rm sc}_{III}(\omega) +\delta g^{\rm sc}_{IV,2}(\omega) \approx 
-\frac{1}{8\pi^2\alpha}
\int\limits_{0}^\infty dx\ x \int \limits_{-\infty}^\infty \frac{dy}{y} 
\Biggl\{ \frac{8 G^{\prime 2}(z)}{G^2(z)} - \frac{4 G^\prime(z) \re G^\prime(z)}{G(z)G(z+i\alpha)}
- \frac{8 G^{\prime 2}(z)}{G(z)G(z-i\alpha)}\notag \\
+  \frac{4 G^\prime(z) \re G^\prime(z)}{G(z)G(z-i\alpha)}
+ \frac{2i\alpha  G^{\prime}(z)G^{\prime\prime}(z)}{G^2(z)}
+i\alpha \frac{2G^{\prime}(z)G^{\prime\prime}(z)+iG^{\prime\prime}(z)\im G^{\prime}(z)+G^{\prime}(z)\re G^{\prime\prime}(z)}{G(z)G(z-i\alpha)}\notag \\
- i\alpha \frac{4[3G^{\prime\prime}(z)+G^{\prime\prime}(z^*)]\re G^{\prime}(z)-G^{\prime\prime}(z)\re G^{\prime}(z)-i G^{\prime}(z) \im G^{\prime\prime}(z)}{G(z^*)G(z-i\alpha)}\Biggr \} .
\end{gather}
Expanding the function $G$ in powers of its argument, we obtain
\begin{gather}
\delta g^{\rm sc}_{III}(\omega) +\delta g^{\rm sc}_{IV,2}(\omega) \approx 
- \frac{\alpha}{\pi^2}\int\limits_{0}^\infty dx\ x \int \limits_{-\infty}^\infty \frac{dy}{y}
\frac{1}{(\bar\epsilon+z)^2(\bar\epsilon+z+i\alpha)(\bar\epsilon+z-i\alpha)}
\notag \\
- \frac{5i \psi^{\prime\prime}(1/2)}{8\pi^2\psi^\prime(1/2)}
\int\limits_{0}^\infty dx\ x \int \limits_{-\infty}^\infty \frac{dy}{y} \frac{1}{(\bar\epsilon+z)^2} = \frac{i\alpha}{6\pi \bar\epsilon^2} W_3\left (\frac{\alpha}{\bar\epsilon}\right ) - \frac{35 \zeta(3)}{\pi^3}\ln \epsilon ,
\end{gather}
where the function $W_3(z)$ is given by Eq.\eqref{eq:W3:def}. Here we neglected terms of the order of $\alpha/\bar\epsilon$.

\end{document}